\documentclass[journal]{IEEEtran}
\usepackage{cite}
\usepackage{hyperref}
\usepackage{amsthm}

\usepackage{color,soul}

\ifCLASSINFOpdf
\usepackage[pdftex]{graphicx}
\else
\fi
\usepackage{amsmath}
\usepackage{tabularx}

\ifCLASSOPTIONcompsoc
  \usepackage[caption=false,font=normalsize,labelfont=sf,textfont=sf]{subfig}
\else
  \usepackage[caption=false,font=footnotesize]{subfig}
\fi

\usepackage{stfloats}
\usepackage{multicol, multirow}
\usepackage{textcomp}

\usepackage{xcolor}

\begin{document}
%
\title{Building Trust in Autonomous Vehicles: Role of Virtual Reality Driving Simulators in HMI Design}
%
%
%

\author{Lia~Morra,~\IEEEmembership{Senior Member,~IEEE,}
        Fabrizio~Lamberti,~\IEEEmembership{Senior Member,~IEEE,}
        F.~Gabriele~Prattic\'o,
        Salvatore~La~Rosa,
        Paolo~Montuschi,~\IEEEmembership{Fellow,~IEEE}
        
\thanks{Copyright (c) 2019 IEEE. Personal use of this material is permitted. However, permission to use this material for any other purposes must be obtained from the IEEE by sending a request to pubs-permissions@ieee.org.} 
        
\thanks{The authors are with the GRAINS -- GRAphics And INtelligent Systems group at the Dipartimento di Automatica e Informatica of Politecnico di Torino, 10129 Torino, Italy. e-mail: (see \href{http://grains.polito.it/people.php}{http://grains.polito.it/people.php}).}
\thanks{Manuscript received XXXX XX, XXXX; revised XXXX XX, XXXX.}}

%
%

\markboth{IEEE Transactions on Vehicular Technology,~Vol.~XX, No.~XX, XXXX~XXXX}%
{Shell \MakeLowercase{\textit{et al.}}: Bare Demo of IEEEtran.cls for IEEE Journals}
%



\maketitle

\begin{abstract}
The investigation of factors contributing at making humans trust Autonomous Vehicles (AVs) will play a fundamental role in the adoption of such technology. The user's ability to form a mental model of the AV, which is crucial to establish trust, depends on effective user-vehicle communication; thus, the importance of Human-Machine Interaction (HMI) is poised to increase. In this work, we propose a methodology to validate the user experience in AVs based on continuous, objective information gathered from physiological signals, while the user is immersed in a Virtual Reality-based driving simulation.  We applied this methodology to the design of a head-up display interface delivering visual cues about the vehicle' sensory and planning systems. Through this approach, we obtained qualitative and quantitative evidence that a complete picture of the vehicle's surrounding, despite the higher cognitive load, is conducive to a less stressful experience. Moreover, after having been exposed to a more informative interface, users involved in the study were also more willing to test a real AV. The proposed methodology could be extended by adjusting the simulation environment, the HMI and/or the vehicle's Artificial Intelligence modules to dig into other aspects of the user experience.

\end{abstract}


%
\IEEEpeerreviewmaketitle

\section{Introduction}
%
%
%
%
\IEEEPARstart{M}{ost} research efforts in the context of intelligent vehicles (IVs) have been directed to improving safety and effectiveness of vehicle's control (autonomy) and vehicle-to-vehicle coordination (connected vehicles) \cite{katoinpress}. To fully reap the benefits of autonomous driving (AD) systems, humans, both drivers/passengers and pedestrians alike, will need to \textit{trust} their safety and reliability. Hence, there is an emerging need to support effective and reassuring communication between humans and IVs. Passengers need to feel confident, at all times, that they have sufficient information about the state of the vehicle, its environment and perceptions as well as its planned and current behavior; even more, that they possess all the appropriate information and means to take over all the aspects regarding the operation of the vehicle in due time, when needed, in a safe and appropriate manner. 

Despite playing a crucial role in the uptake of any system based on autonomous agents, including autonomous vehicles (AVs), trust between humans and machines is generally hard to establish. According to a 2017 survey by the Pew Research Center on ``Automation in everyday life'', over half (56\%) of the Americans who were interviewed said they would not want to ride in a driverless vehicle if given the opportunity \cite{smithanderson2017}. 

However, preliminary experiments in the literature on partially autonomous driving scenarios show that these negative emotions can be reduced by adopting Human-Machine Interaction (HMI) designs that provide feedback about how the car is acting (what automated activity it is undertaking) and the reasons why the car is acting that way \cite{koo2015did}. 

The role of HMI in IVs is thus profound and, for this reason, user experience (UX) should be taken into large account at any stage of the development process. By establishing a collaborative relationship between drivers/passengers and vehicles, HMI can positively affect the acceptance as well as the technological advancement of AD solutions. 

Unfortunately, the application of consolidated approaches for UX design and evaluation to AD systems is not straightforward. For instance, focusing on the quantitative assessment of a particular user interface design, techniques that measure driver's performance in specific driving tasks could not be easily reused when, due to the specific level of automation, there are no more drivers but passengers. Similarly, post-experience questionnaires (alone) could be no more appropriate when feedback to be collected concerns the huge amount of aspects that may contribute to the perceived level of trust. Even driving simulators that are used today for developing vehicle's intelligent behaviors may not be directly applied to UX studies, as focus would have to be shifted, e.g., on the vehicle's interior and on interaction with it, rather than on the fidelity of external factors affecting its decisions (traffic, presence of pedestrians, etc.). 

By moving from the above considerations, in this paper we present a methodology that is meant to support the study of HMI with IVs, and we show its helpfulness in the evaluation of the passengers' level of trust by considering the design of a possible interface for AD systems. 

The devised methodology relies on a simulation platform based on immersive Virtual Reality (VR), which was developed by grounding on an existing driving simulator. Although, in principle, the technology is applicable to many scenarios, from unassisted to fully autonomous systems, we focused on L4 and L5 automation levels, as they represent the configurations for which characterizing the passenger's experience from the point of view of comfort and trust is more challenging. We therefore created a virtual AD system that allows users to experience a simulated ride in a virtual urban environment, facing a number of different situations.   

For the assessment of the UX, we consider both cognitive and affective factors, by integrating feedback based on subjective post-experience questionnaires with continuous, objective information gathered  from physiological signals. In particular, in this paper we focused on stress level measurements to investigate the perceived degree of safety and ``connection'' with the vehicle. Notwithstanding, the proposed methodology has been designed in a way to support later extensions for the detection of other emotional states. It is worth observing that, thanks to its immersive nature, VR allows to measure the latter state much more realistically than other traditional simulation scenarios \cite{eudave2017physiological}.

With the aim to evaluate the suitability of the proposed approach, the methodology was applied to the design of a head-up display (HUD)-based interface for AVs that provides visual cues about the vehicle' sensory and planning systems. As said, providing information about how and why the car is acting is crucial to elicit trust in AVs, but little experimental evidence is available to determine how such information is best presented to the passengers \cite{Jose, Lungaro}. By applying our approach to the above scenario, we obtained qualitative and quantitative proofs that a complete picture of the vehicle's surrounding, despite the higher cognitive load, is conducive to a less stressful experience. Moreover, after having been exposed to an interface delivering a higher information content, users involved in the study were also more willing to test a real AD system. 

Besides offering interesting insights that may drive future HMI designs, the results confirm the effectiveness of the proposed methodology in digging into a use case that well represents possible facets of the UX which could be investigated through the experimented techniques.

\section{Background and Related Work}
\subsection{HMI in Partially and Fully Automated Vehicles}
\label{sec:background-hmi}
Establishing trust is important in order for users to accept, and even rely on, automated systems. Mcknight \& Chervany \cite{10.1007/3-540-45547-7_3} have identified three constructs necessary to increase trust: ability, benevolence, and integrity. When the trustee is an autonomous system, these factors translate in the system's \textit{performance} and skillful execution, into the sharing of a common \textit{purpose} with the user, and into the implementation of a reliable and consistent \textit{process}. Trust is thus established through direct observation of the system's behavior and its underlying mechanisms. Lee \& See observed that ``Trust that is based on an understanding of the motives of the agent will be less fragile than trust that is based on the principle of reliability of the agent'' \cite{LeeSee}. In the context of AD systems, HMI plays a fundamental role in this respect, by providing information about the vehicle's performance. In fact, partially automated vehicles on the market allow the driver to monitor the status of the car's components. User interfaces are designed to increase the perceived ability of the system and to support predictability, thus inducing trust. 


In recent years, a study by Ekman and colleagues provided a systematic review of HMI design principles that promote trust in AD systems  \cite{ekman2018creating}. The authors distinguish a \textit{learning phase}, that starts with the first interaction and lasts until the user is familiar with the AD systems, from the \textit{performance phase}, which takes into account a long-term use perspective. During a testing simulation, it can be argued that the learning phase is most important, although its specific duration differs on an individual basis. In the performance phase, trust is mainly based on the performance and dependability of the system, and is fairly stable unless an error or unexpected event occurs; in the learning phase, it is the user's ability to form a mental model of the AD systems that is crucial to form a trust bond. 

Hence, in this work we focused our attention specifically on the four factors that, according to \cite{ekman2018creating}, are more relevant for the learning phase: \textit{mental model}, the ability to form an approximate representation of the AD system's skills and functions; the system's proneness to be perceived as an \textit{expert}/\textit{reputable} agent; the possibility to provide continuous \textit{feedback} to the user, ideally addressing two or more senses; finally, the provision of \textit{how and why information} regarding upcoming actions. In this context, a ``how'' message describes how the system solves a given task, whereas a ``why'' message pertains to the motivations that lead to the task itself. 

A limited number of experimental studies have, so far, established that providing information to the driver/user usually increases driving performance and acceptability in partially  \cite{verberne, Hauslschmid, koo2015did} and fully automated driving systems \cite{Lungaro}. For instance, Verberne et al. \cite{verberne} found that Adaptive Cruise Control (ACC) systems that share the same drivers' objectives, like the adoption of a relaxed and safe driving style without sudden braking and accelerations, while at the same time providing information to the user, are considered more reliable and acceptable. Koo et al. \cite{koo2015did} explored the effect of providing ``how'' and ``why'' information in the context of an auto-braking system. Providing both the information types resulted in the safest driving behavior, at the expense, however, of a high cognitive load and decreased acceptability. Drivers preferred receiving only ``why'' information, whereas the ``how'' information was often perceived as redundant. The interfaces considered in these studies were very simple compared to the technical possibilities of current user interfaces: they consisted of brief verbal messages, with no visual cues \cite{koo2015did}, or included only information on the position of obstacles \cite{Lungaro}. 

It is important to consider not only which information is provided to the users, but also how it is conveyed. The visual mode is the primary and most widely used among the vehicle interfaces, and represents the most consistent communication channel. In-vehicle display devices can be grouped in three categories: \textit{head-down displays} (HDDs), \textit{head-up displays} (HUDs), and \textit{head-mounted displays} (HMDs). HDDs offer the advantage of not blocking the view of the real world for the users, who, however, find themselves distracting from the road. HUDs make it possible to take advantage of the necessary information while keeping an eye on the external environment, but pose significant construction challenges. HMDs share the advantages of HUDs, but only a few devices are available on the market, which suffer from some usability issues (especially for in-vehicle applications). 

Studies have consistently shown that HUDs result in a better driving experience and performance than HDDs, leading to shorter reaction times \cite {doshi2009novel}, decreased cognitive load \cite{medenica2011augmented}, and fewer driving errors \cite{KimDey, Jose}; HUDs are also preferred by users against both HDDs and HMDs \cite{Jose}.  Augmented Reality HUDs (AR-HUDs) have been found especially effective in  increasing the driver's intuitive cognition \cite{ParkAR} and promoting a safer and more effective driving behavior, particularly in demanding driving situations \cite{HaeuslschmidcontactAnalog,7795724}. 

Given the technical difficulties in realizing AR-HUDs, current displays often come in the form of prototypes, concepts or demonstration videos. Examples in the literature often focus on specific aspects of the driving experience, such as driver assistance (DA) \cite{Jose} and obstacle detection \cite{Lungaro}. Many commercial prototypes focus on partially automated systems that extend current DA solutions, whereas Waymo and NVIDIA are more directly focused on L4 and L5 automation. 

Information displayed by the main commercial solutions is reported in Table \ref{tab:HUD_concepts}. A tendency to adopt a common set of symbols and metaphors can be observed among vendors. For instance, information related to ACC and Lane Keep Assistance functionalities, such as the current lane, speed, and the position and speed of preceding cars are displayed by Continental, Hyundai, PSA, and Daqri. Waymo and NVIDIA include richer information on both the path planning and the sensory capabilities of the vehicle. Through bounding boxes (i.e., parallelepipeds enclosing detected objects), colored overlays and other elements, all the factors involved in driving are highlighted. In addition, navigation information is added not only for the user's vehicle, but also related to other cars, pedestrians or cyclists through motion prediction.

\begin{table}
	\caption{Information displayed by commercial HUD concepts and demonstration videos. }	
	
	\begin{tabular}{p{3cm} p{5cm}}
	\hline
	AR-HUD & Displayed information \\
	\hline
	Continental AR-HUD Concept & Lane Departure Warning System (LDWS), Assisted Navigation, Adaptive Cruise Control (ACC) \\
	Hyudai  AR-HUD Concept & Traffic lights, Assisted Navigation, LKA, ACC \\
	PSA group  AR-HUD concept & Assisted Navigation, LKA, ACC, Pedestrians, Approaching obstacle warning  \\
	Daqri  AR-HUD Concept & Assisted Navigation, Lane Keep Assistance (LKA), Lane Control, ACC, Approaching obstacle warning, Children crossing, Pedestrians \\
	WayRay Holographic AR Display concept &  Assisted Navigation  \\
	Waymo Demo video & Traffic signs, Cars, Pedestrians, Cyclists (bounding boxes and colored overlays with distance and speed information), Motion Prediction, Assisted Navigation  \\ 
	NVIDIA Drive AGX &  Traffic signs, Cars, Traffic lights, Lanes, Pedestrians, Cyclists (bounding boxes with distance information), Motion Prediction, Lane separation lines, Route planning data \\
	\hline
	
	\end{tabular}

  \label{tab:HUD_concepts}
\end{table}

\subsection{Measuring User Experience in Driving Simulators}
\label{sec:background-biosig}

Researchers have for long time relied on driving simulators to cope with difficulties and risks associated with field testing \cite{guo2018automatic}. In recent years, VR simulators have elicited a lot of interest thanks to their immersive nature \cite{Jose,medenica2011augmented, FabrizioVRSim, chen2018driver}. 

Most studies investigating different aspects of driving in simulated scenarios, including HMI design \cite{wang, Jose}, rely on drivers' behavior and performance as a proxy for their emotional and cognitive status \cite{fisher2011, koo2015did, doshi2009novel}. Experimental measures include standardized questionnaires as well as indicators such as driving speed, lane keeping, braking patterns, etc., for which absolute or relative validity has been generally established \cite{fisher2011}. However, in AD systems, humans are expected to take progressively less part in driving, which makes behavioral assessment less relevant.  

Physiological signals are increasingly used to measure users' affective and cognitive states in engineering in general \cite{balters2017capturing}. The activity of the autonomic nervous system, which regulates affective states, can be captured non-invasively through signals such as Heart Rate (HR) and Electrocardiography (ECG), Electromyography (EMG), Respiratory Rate, and Galvanic Skin Response (GSR). In the last years, researchers also investigated their use combined with traditional or immersive driving simulators \cite{fisher2011,ruscio2017collection,eudave2017physiological}.  

In particular, the relative validity of physiological signals for traditional driving simulators is supported by several studies, albeit available data is less abundant than for driving performance \cite{fisher2011, johnson2011physiological, eudave2017physiological}. For instance, risk perception was found to be highly correlated with changes in GSR \cite{fisher2011}. Comparison between on-road and simulated driving conditions established the relative validity for mean HR and mean oxygen consumption, although HR values observed in real driving conditions were higher, probably due to the increased stress associated with driving on a real road  \cite{johnson2011physiological}. In a pilot study, Eudeave and colleagues found that the physiological response in an immersive VR environment is stronger than in a traditional driving simulator  \cite{eudave2017physiological}. 

Recording of physiological signals have also been exploited in real-life driving conditions to characterize drivers' performance and experience, from measuring stress levels to detecting drivers' drowsiness  \cite{healey2005detecting, singh2014assessment}. Of particular interest is the study of Healey and colleagues on driving-related stress \cite{healey2005detecting}. ECG, EMG and GSR were recorded while drivers followed a set route; driving sessions were videotaped and visually inspected for observable stress-induced actions, such as head turning, to be used as reference standard. Collected signals allowed the authors to distinguish different levels of stress with high accuracy (over 97\% across multiple drivers); GSR and HR metrics were most closely correlated with drivers' stress level. Again, studies have been conducted, so far, from the point of view of an active driver, leaving the question open on whether stress-induced changes can be equally and as effectively observed in passengers.

\section{Proposed Methodology}

\subsection{Overview}
\begin{figure*}[t]
    \centering
    \includegraphics[width=1.55\columnwidth]{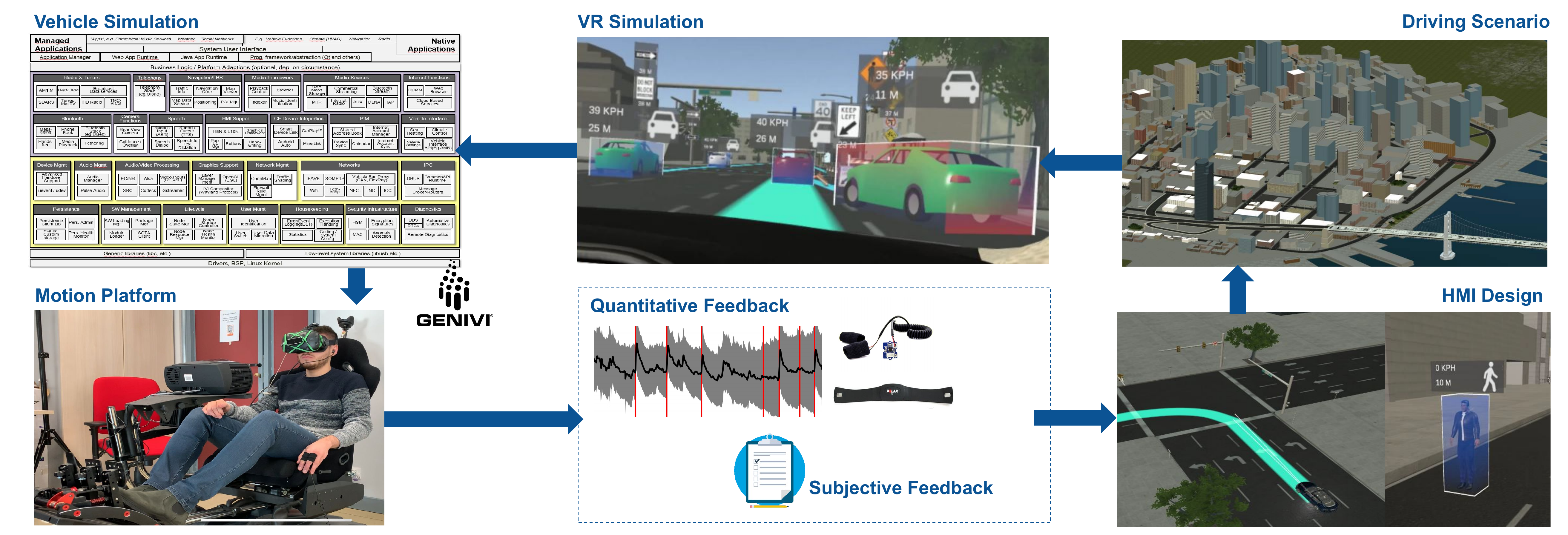}
    \caption{Proposed methodology: exemplification on the task of HMI design. A defined AV driving scenario is simulated in immersive VR. Vehicle simulation is based on the open source GENIVI platform. The simulator is integrated with a motion platform to further foster immersion. User's feedback is collected through both subjective, offline questionnaires, and objective, real-time physiologic measurements reflecting cognitive and affective states.      }
    \label{fig:architecture}
\end{figure*}

As discussed in Section \ref{sec:background-hmi}, trust in automated systems can be achieved from direct observation of system's behavior, coupled with an understanding of the underlying mechanisms. To this aim, as depicted in Fig. \ref{fig:architecture}, the devised methodology relies on a VR-based AV simulator. Simulation allows the user to get immersed in repeatable scenarios including a variety of both ordinary and emotional-intensive events. User is provided with insights on the autonomous system's behavior by means of a virtual AR-HUD combined with additional audio cues. In this way, we postulate that the user can form an adequate mental model of the AD system. Assessment is performed by collecting feedback from the user in the form of subjective (questionnaire-based) ratings and objective (physiological signal-based) measurements.

\subsection{Technology and Setup}
The VR system is implemented using the HTC Vive eco-system (https://www.vive.com/eu/product/), by HTC Corporation, Taiwan. The Vive VR Headset features a resolution of 1080$\times$1200 pixels per eye spanning a horizontal 110$^{\circ}$ FOV at 90Hz. The native positional tracking leverages the IR lasers emitted from the Vive Base stations (built upon the Valve's Lighthouse technology) which, combined with headset's built-in sensors, enables a 6 DOF \textit{outside-in} tracking of the user's head.

With the aim to foster immersion through the simulation of the motion stimuli that a driver or passenger would experience on a real vehicle, an inertial motion platform is used.  The platform exploited in this work is the Atomic A3 Racing, designed by Atomic Motion Systems, which supports 2 DOFs (yaw and pitch) motion simulation. To simulate the user's perceived accelerations, the so-called \emph{tilt coordination} motion simulation strategy [7] was implemented. In short, this technique works by imitating the perceived acceleration via decomposition of the gravity acceleration vector, obtained through a coherent rotation of the platform. A motion compensation needs to be applied to the VR coordinate system (which is centered in the headset, i.e., in the user's viewpoint), based on current platform's rotation. To this purpose, a Vive Tracker was mounted on the seat and tracked together with the headset. 

Finally, since it has been proved that letting the user see his or her hands in the virtual environment  increases the sense of presence \cite{dalgarno2010learning}, a virtual replica of the user's hands including articulated fingers is created by tracking them using a Leap Motion Controller device attached to the headset.

\subsection{AV Driving Simulator}

The vehicle simulator implemented is based on the open source Simulator Vehicle project by the GENIVI Alliance \cite{genivi_details} (in the following simply referred to as GENIVI, for brevity). GENIVI was selected among several possible alternatives for multiple reasons: it was originally created to support HMI design; it allows, by design, the addition of new features; it already provides modules for intelligent traffic simulation; it includes a basic \emph{auto drive} functionality for the user's vehicle; it provides a few driving scenes and vehicles with their own rigid body physics-based controller. 

The main activities carried out to adapt GENIVI to the purposes of this work involved: porting of available features to VR; integration of the motion platform; implementation of a custom AD controller. The latter activity was considered as necessary since, in a preliminary study, the built-in controller was judged not realistic enough, especially when dealing with complex, unpredictable events (e.g., sudden pedestrian crossing, etc.).

\subsubsection{VR porting}
Implementing the support for VR was facilitated by the fact that GENIVI is based on the Unity game engine, which natively allows for the creation of VR applications for the HTC Vive. Our implementation allows to virtual accommodate the user to any seat of the virtual vehicle. Built-in vehicles, namely a Land Rover L405 and a Jaguar XJ, are designed for a non-immersive simulation. Hence, a new vehicle was created with VR-based interaction in mind, i.e., by focusing on visual fidelity of the vehicle's interior. Finally, support for users' virtual hands was added. 

\subsubsection{Motion platform integration}
to integrate the motion platform, an additional software module was developed. The module receives in input the acceleration values calculated in the seat's tracked point by the physics simulation engine and outputs it to the proprietary platform's driver (AMS Symphinity), which remaps them to coherent tilt and pitch angles and consequently applies them to the platform. Other motion platforms may be integrated in a similar way.

\subsubsection{AV simulation}
within this work, our aim was to provide a methodology to study the considered domain using simulated VR-based scenarios, accompanied by suitable measurement tools, rather than contribute to the advancement of the state of the art of AVs' control sub-systems. Basic AD functionality available in GENIVI was therefore extended to make it cope with situations of interest. Attention was focused on reproducibility, while preserving simplicity. More sophisticated implementations, leveraging, e.g., data provided by vehicle's virtual sensors could nonetheless by integrated in the future.

Our implementation takes advantage of the native trajectory system, which is used in GENIVI to manage the traffic. Paths to follow are embedded in the scene description using a complex network of waypoints. The developed AD system relies on it to feed a PID-based controller, which is in charge of driving the vehicle by making it accelerate, brake, and steer. Differently than with the other cars in the traffic, the AD system is affected by the full set of accurate, rigid body physics simulation variables. The PID was fine-tuned in closed loop using manual parameter adjustment targeting a maximum overshooting of 5\% at step response, in order to achieve a comfortable and realistic behavior. To this aim, control commands shaping and auxiliary waypoints were also used. Although different and far more sophisticated approaches could be investigated in the future (e.g., \cite{nestedpid}), the selected control system proved to meet the simplicity-effectiveness trade-off required to cope with the issues tackled in this work. Appropriateness of the pursued approach was also confirmed by subjective observations concerning simulation quality (Section \ref{sec:questionnaire_results} and Supplemental material).

The same approach was pursued also to bind specific vehicle's reactions to the pre-programmed events. Obstacle avoidance is handled by a dedicated logic, which also takes into account trajectory replanning when the obstacle cannot be avoided by simply adjusting vehicle's speed. For moving obstacles, replanning takes into account predicted motion. Further details on the implementation can be found in \cite{tesi_laurino}.

\subsection{HUD Design}
\label{sec:HUD}

Based on the principles discussed in Section \ref{sec:background-hmi}, an in-vehicle user interface should continuously provide feedback addressing, whenever possible multiple senses, highlighting ``why'' information that explains the vehicle's choices, and adopting a pleasant and effective communication style that presents the system as a skilled/reputable driver. These elements, while important in general, are particularly relevant in the initial learning phase, where the user is still unfamiliar with the AD system and needs to form an appropriate mental model of its inner mechanisms \cite{ekman2018creating}. As it will be illustrated in more detail in Section \ref{sec:experiments}, subjects who participated in our study were never exposed to a real AD system. An AR-HUD was therefore designed, as it was found in the literature to be the most effective interface under the considered conditions. 
 
It was deemed as important to ensure that visual cues displayed by the AR-HUD are consistent with information conveyed by commercial DA products, as users are mostly familiar with it. However, it was regarded as crucial to provide also information that illustrate the vehicle's sensory capabilities and hence, improve the user's situational awareness. Finally, given this work's focus on L4 and L5 AD systems, information about the vehicle's planning functionalities needed to be delivered as well. 

Design was based on the features reported in Table \ref{tab:HUD_concepts}. The HUD is capable to display information about all the relevant elements in the surrounding environment, including both static objects (trees, lighting poles, parked cars, traffic lights, road signs, etc.) and dynamic objects (pedestrians, animals or other cars). These elements are provided together with distance information in meters from the vehicle, absolute speed when available, and a visual warning status indicator. Lane keeping and navigation cues for the user's vehicle and other cars (assuming that they are connected) are also considered. The color of each car is randomly assigned by GENIVI.

Objects of interests are identified by means of a bounding box. This metaphor, previously validated in the literature \cite{7795724}, is adopted by commercial players such as Waymo and NVIDIA. In our implementation, each bounding box has a white outline and is associated with a label and an icon identifying the detected object, thus satisfying the usability principle which suggests that the adopted representation must be simple and intuitively understandable by the user \cite{hancock}; the use of familiar cues, such as icons, also reduces the cognitive load in the presence of a large amount of information \cite{Dekker}. Bounding boxes are automatically generated in VR knowing the position, size and pose of all objects in the scene. Technically, this was implemented in Unity by associating to all objects with a \textit{Collider} component a visible colored material. The \textit{Colliders} are not visible by default in the rendering step because just define the bounding volume of an object for the purposes of identifying physical collisions through the physics engine. Labels always face the vehicle and are, therefore, readable by the user.  

In order to determine which situations constitute a potential danger, we relied on the definition from the ISO 15623 standard on ``Forward vehicle collision warning systems'' \cite{dangerEvaluation}, counting on a previous study by Sebastian et al \cite{5164404}. A mathematical model is used to determine potential collisions based on the trajectory, speed and acceleration of the vehicle, as well as that of potential obstacles (e.g., the preceding car). Once the possibility of a collision has been established, a safety distance is calculated, which depends on the speed of the vehicle and the reaction time of the driver, which was estimated based on the study in \cite{reactionTime}. The distance between the vehicle and the estimated collision point is therefore measured: if this distance is less than the safety distance, the passenger needs to be warned of the potential danger. We therefore defined a \textit{hazard index}, ranging from 0 to 1 and calculated as the ratio between the distance from the obstacle and the warning distance defined in \cite{5164404}. 

The objects' warning status is presented through both visual and auditory cues. In the literature, the AR-based DA system in \cite{6232220} adopted an intuitive color code in which the severity of the danger of an obstacle detected on the road is shown by means of a color code that starts from the green (safety) and extends up to the red (maximum state of danger). This color coding is consistent with systems reviewed in Section \ref{sec:background-hmi}. Therefore, we decided to color-code the hazard index with a green to red gradient and use it to visually represent the warning status of the detected object by controlling the transparency color of the bounding box. The color-code value is computed using a perception-based equation, where the hazard index is used as exponential factor.

To signal potential dangers, the label associated with the bounding box flashes as well, so as to direct the user's attention towards the obstacle. Flashing is used in DA systems by various vendors \cite{bmwPed, volvoCWAB}. It is important to underline that, through the flashing information, the vehicle communicates to the user why it is about to perform a specific action \cite{ekman2018creating}. The flashing visual cue, at a lower frequency, is also used to notify the user of a road sign or traffic light. Flashing occurs when a traffic light changes or when a new road sign is recognized. The lower frequency reduces the sense of alarm and, hence, allows the user to distinguish normal driving operations from high-risk situations. 

An immediate danger is also marked by a sound alert \cite{ferati2017universal, dangerEvaluation}. This is consistent with current DA systems, which produce audible warnings, e.g., in emergency braking conditions \cite{volvoCWAB}. A more pleasant sound is played when road signs are detected, to capture the user's attention in an unalarming way. 

Two variants of the HUD were designed, which in the following are referred to as \textit{omni-comprehensive} (OMN) and \textit{selective} (SEL). 

\subsubsection{Omni-comprehensive HUD}
in the OMN variant, we show information about all dynamic elements (cars and pedestrians) within a ``detection'' diameter which is set to 150 meters. This threshold was firstly motivated by practical reasons: virtual objects beyond this distance would be too small to be appreciated considering the resolution of the display in the VR headset. This distance is also compatible with the equipment of current AD prototypes and the detection range of LiDAR systems. Road signs and traffic lights are always shown in the interface, except for those that regulate road sections different than the one the vehicle is currently on. Furthermore, it was decided to exclude from the display the information about static objects such as trees, parked cars, lighting poles, etc. unless they become dangerous. This exclusion is motivated by the principle of cognitive load, according to which an interface should be easily understandable by the user, simple and intuitive, as it avoids excessive cluttering \cite{hancock}.

\begin{figure}
    \centering
    \subfloat[OMN]{\includegraphics[width=0.8\columnwidth]{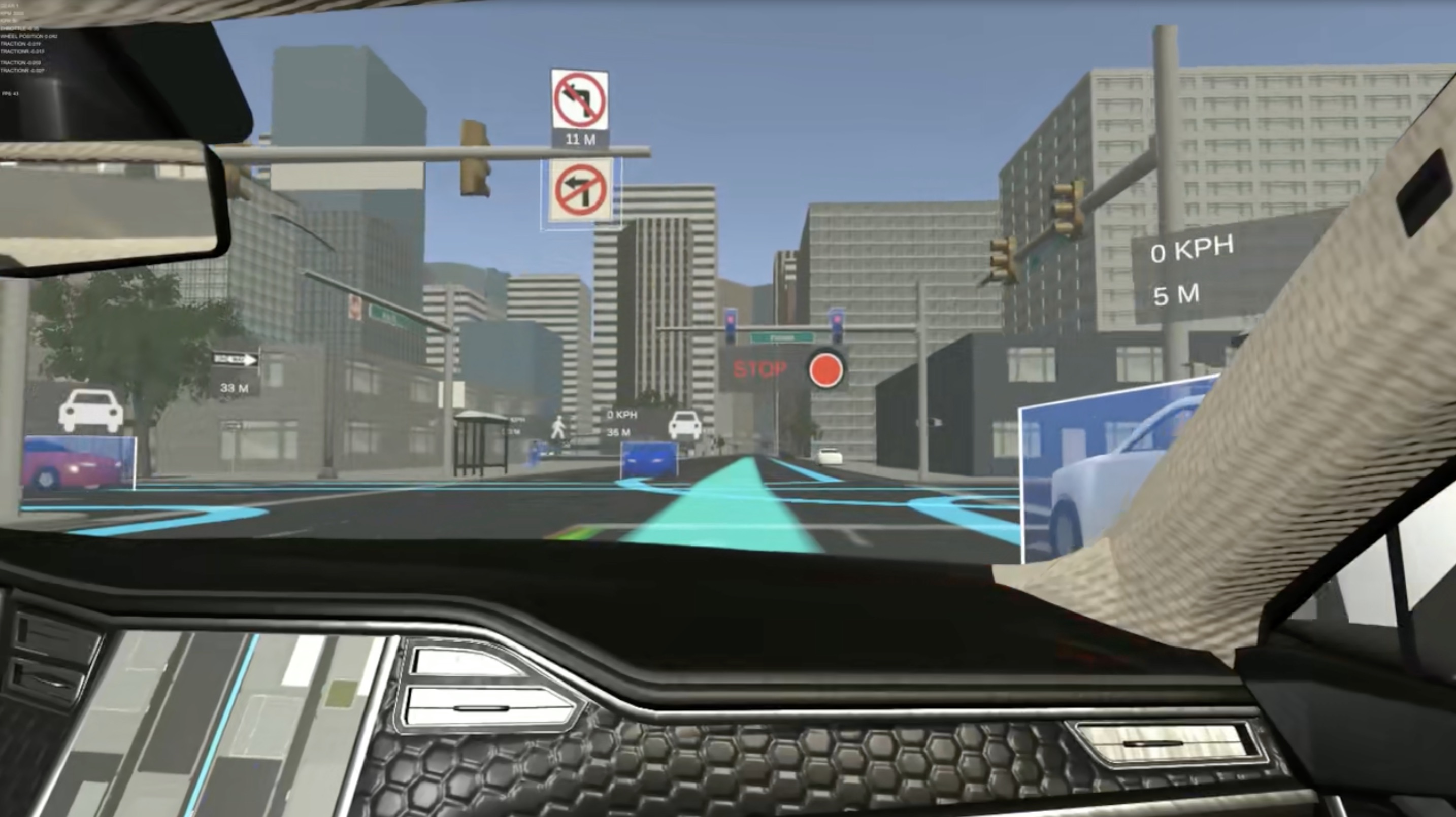}}
    
    \subfloat[SEL]{\includegraphics[width=0.8\columnwidth]{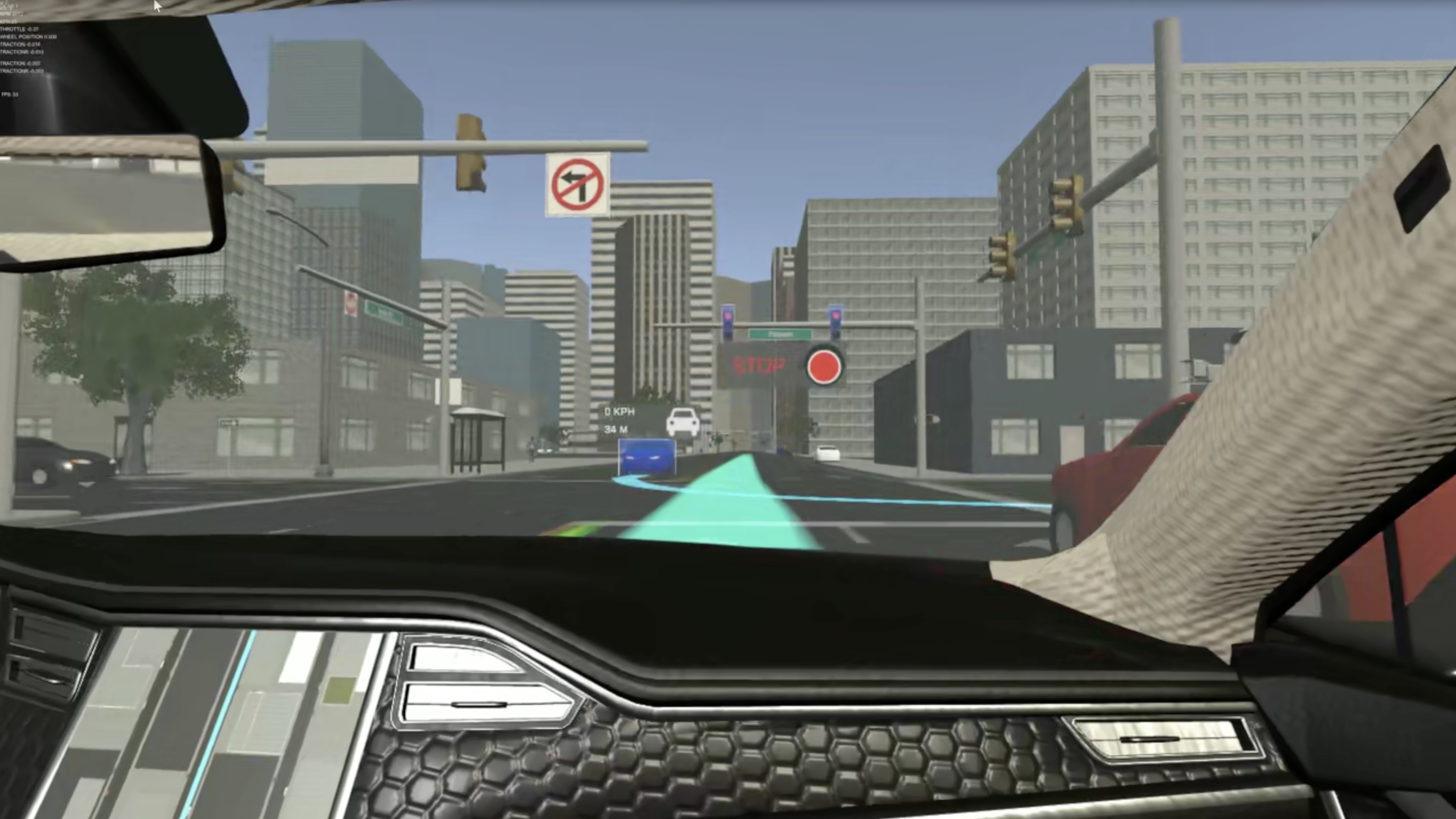}}

    \caption{Comparison between the OMN (a) and SEL (b) AR-HUD interfaces.}
    \label{fig:hud_comp}
\end{figure}

\subsubsection{Selective HUD}

in the SEL variant, only information that is deemed of specific interest to the user is displayed. The guiding principle was to select information that pertains to those elements of the environment that, at any given point, affect the behavior of the AD system. Let us consider road signs: the vehicle detects all road signs within the diameter of interest, but not all of them are necessarily useful at the time. For example, in the presence of a pedestrian crossing sign and a speed limit sign, the vehicle may decide not to show any information on the former sign, based on the fact that, at the moment, there is no pedestrian intending to cross;  the latter sign may  force the vehicle to slow down and, thus, it would be highlighted in the interface. 
More specifically, in the SEL variant only cars that precede the current vehicle or, more generally, that intersect its current trajectory, are highlighted with a bounding box. Pedestrians and other static or moving objects are identified only if and when they become dangerous, i.e., when a collision becomes possible. Navigation lines for other cars are only displayed when assessed by the vehicle (e.g., at intersections to determine priority). Traffic lights information, as well as tracing of the vehicle's navigation line and the road center line, are unchanged in this variant. A comparison of the two interfaces is provided in Fig. \ref{fig:hud_comp}.

\subsection{Simulated Scenario}

In order to create a relationship of trust between a user and an AD system, the latter must show its ability in dealing with different driving scenarios \cite{LeeSee}. Our simulated scenario is constructed to include a variety of different situations, both ordinary and challenging, that may occur in an urban setting. The urban setting is, in general, considered the most difficult to manage by AVs \cite{kim2015_urban_complexity}. In fact, current L2 and L3 automated systems are mostly restricted to motorways and extra-urban routes, and the biggest challenge in the development of L4 and L5 systems is represented precisely by urban areas where significantly more factors are at play, driving conditions are far less predictable, and the presence of pedestrians amplifies the perceived risk. 

Compared to real-life driving, simulators offer the distinctive advantage of creating repeatable scenarios, where most experimental factors can be easily controlled. Therefore, it is possible to study and compare subjects' reaction to individual events, whereas in real-life driving experiences one would be mostly restricted to consider overall measures \cite{healey2005detecting, johnson2011physiological}.

The simulation was created starting from one of the scenes included in the GENIVI platform, representing a miniature version of the city of San Francisco. As said, despite the basic auto drive functionality, GENIVI is natively meant to support mostly first-person driving experiences, with random traffic patterns, no pedestrians and no intentionally hazardous situations. By leveraging the developed AD capabilities and the integrated waypoint system, different situations were embedded in the simulated scenario in order to showcase different AD abilities and elicit changes in the subjects' affective state. Considered abilities include: interacting with traffic and especially with other cars, e.g., maintaining safety distance, overtaking, etc.; handling road signs and traffic lights; avoiding obstacles and dealing with other potentially dangerous situations, including those where other cars or pedestrians do not behave correctly. 

\begin{figure}
 \centering
 
 \includegraphics[width=6cm]{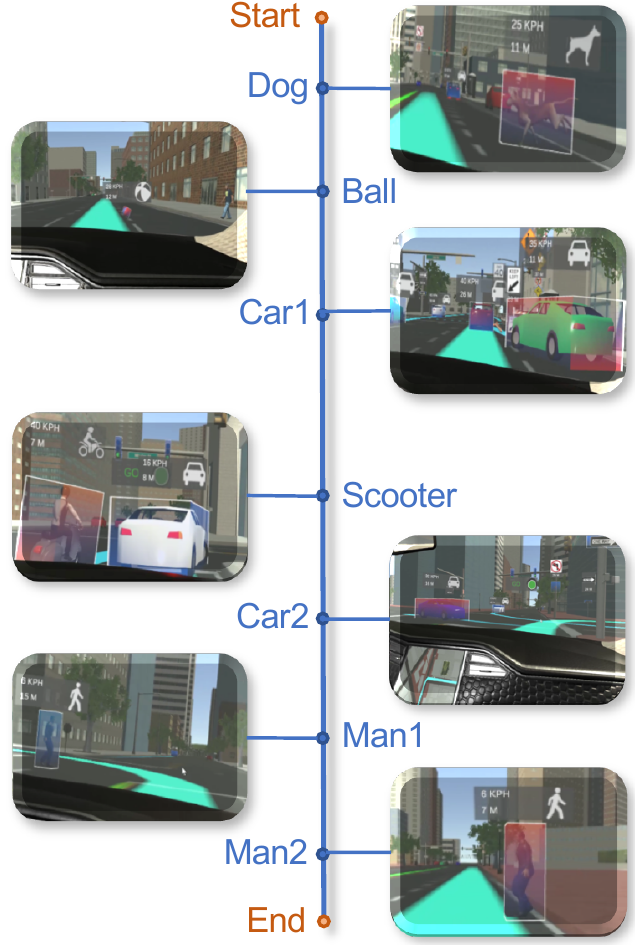}

\caption{Timeline of the test scenario with simulated events.}
\label{fig:test_situations}
\end{figure}

The subject is seated in the passenger's position (front right). The experience begins in an area with a relatively simple environment and little traffic, to allow the user to familiarize with the AD system. Then, the environment becomes populated by cars and pedestrians: the subject becomes familiar with the HUD and the way information is conveyed by the vehicle. Afterwards, riskier situations occur in which the vehicle can show its decision-making skills. In \cite{LeeSee}, it was observed that ``If trust is primarily based on rules that characterize performance during normal situations, then abnormal situations might lead to the collapse of trust''. This strengthens the importance of including driving situations that, while less likely, may pose significant challenges for an AD system. To simulate a typical urban context, such situations were spaced throughout the simulation and alternated with ordinary ones, as illustrated in Fig. \ref{fig:test_situations}. After every risky situation, the car stops for few seconds, to ensure that the subject has enough time to understand what happened and to reflect on how the car handled that situation. Considering the time for letting subjects get acquainted with the system as well as the time required to achieve a suitable distribution of situations, the duration of the simulated scenario was set to 12 minutes. 

Simulated events include the sudden crossing of a dog (Dog), a child on the sidewalk throwing a ball on the street (Ball), scooters and cars that split lanes while driving (Scooter, Car1 and Car2), as well as pedestrians crossing the street (Man1 and Man2). Illustrative frames are reported in Fig. \ref{fig:test_situations}. 
The Dog event corresponds to a highly hazardous situation, in which the vehicle is forced not only to slow down, but also to steer in the opposite direction to avoid a collision. The same happens in the Ball and in the Man2 events (in the latter case, a pedestrian crosses outside of a designated crosswalk while the car is at full speed). Man1 is a less risky situation, as the car is approaching a red light and is already braking when the pedestrian starts crossing. In the Scooter event, the vehicle slows down as the preceding car turns right; in the meanwhile, a scooter enters the lane from the left. The situation is not particularly dangerous, as the vehicle was already reducing its speed to deal with traffic jam; however, from the viewpoint of vehicle-to-human communication, this interaction is complicated as it involves several vehicles. In the Car1 event, a car suddenly changes its lane when approaching road construction (which is poorly visible), forcing the vehicle to quickly reduce its speed to avoid a collision. The Car2 event is even riskier, as another car driving on the intersecting road does not stop at a red traffic light and instead passes at full speed, forcing the vehicle to brake very abruptly. 

Two videos showing the simulated scenario with the OMN and the SEL interfaces are available at http://tiny.cc/p4v16y.

\subsection{Galvanic Skin Response}

As discussed in Section \ref{sec:background-biosig}, physiological signals related to the activity of the autonomic nervous system can provide non-invasive information about the user's affective state. While, in principle, a combination of different signals can be used, for the sake of simplicity in this work we focus on the GSR signal, which was found to effectively detect stress in both simulated and real-life driving \cite{healey2005detecting, fisher2011}. Furthermore, it is easily measured with a simple sensor placed on the fingers \cite{balters2017capturing}.  GSR is mostly sensitive to the dimension of \textit{arousal}, going from sleepiness to excitement or stress \cite{valenza2012role}; it leaves open whether the arousal change is of a positive or negative nature (the \textit{valence} dimension), which nonetheless in our specific case can be derived from the context.

The GSR can be decomposed into a slowly changing tonic component, the Skin Conduction Level (SCL), and an impulsive phasic component,  the Skin Conductance Response (SCR) \cite{valenza2012role}. While the SCL reflects the overall emotional state as well as habituation to the environment, the SCR measures activation in response to a stimulus, e.g., a potentially stressful event occurring in the simulated scenario. The magnitude of the response should correlate to the perceived threat. This phenomenon was previously validated in other types of VR environments \cite{meehan2002physiological}, with an observable effect on the GSR signal even after multiple exposures. 

\subsubsection{Signal processing}
the SCR data was extracted using a 3\textsuperscript{rd} order Butterworth band-pass filter ranging from 0.16 Hz to 2.1 Hz \cite{singh2014assessment, valenza2012role}. Normalization is required to account for the intrinsic inter-individual differences in skin conductance \cite{balters2017capturing}. The most common choices are \textit{z-score} standardization, in which the signal is divided by the standard deviation after subtracting the mean, and \textit{min-max} normalization, in which the signal is normalized between 0 and 1. We found that the min-max scaled signal was most useful for visualization purposes and trend analysis, whereas z-score normalization could be used for the final feature extraction as we observed less inter-subject variability. Considering the fact that, typically, SCR peaks appear between 1 and 5 seconds from the stimulus's onset and last for about 10 seconds, we extracted the SCR waveform for a time window of \textpm 10s centered on each event \cite{figner2011using, slater2006analysis}. Within each window, all samples are divided by the initial signal value, to focus on relative changes.

\subsubsection{Feature extraction}
\label{sec:feature}

for each event time window, a set of features is extracted \cite{figner2011using}. Let $\hat{GSR}(k,j,i)$ be the z-score standardized data value for subject $k$, event $j$ and sample at time $i$, $SCR(k,j,i)$ the corresponding filtered signal representing the skin conductance response, and $L$ the total number of samples per time window.  Features extracted include the mean GSR (Eq. \ref{eq:GSR_mean}), the accumulated GSR (Eq. \ref{eq:GSR_Acc}), the max GSR (Eq. \ref{eq:GSR_Max}, and the Peak to Peak distance in SCR (Eq. \ref{eq:GSR_Peak}). Each feature is calculated on the 10s before ($Pre$) and the 10s after ($Post$) the test event, and the difference ($\Delta$) is used as the final measure.
\begin{gather}
\overline{GSR}_{mean}(k,j)=\frac{\sum_{i=0}^{L}\hat{GSR}(k,j,i)}{L} \label{eq:GSR_mean} \\
{GSR}_{Acc}(k,j)=\sum_{i=0}^{L}\hat{GSR}(k,j,i) \label{eq:GSR_Acc}\\
{Max}(k,j)=\max_i\big(\hat{GSR}(k,j,i)\big) \label{eq:GSR_Max}\\
P2P(k,j)=\max_i\big(SCR(k,j,i)\big)- \min_i\big(SCR(k,j,i)\big)
\label{eq:GSR_Peak}
\end{gather}

\subsection{Questionnaire}
Subjective data about the experience can be collected through questionnaires. The questionnaire that we designed tackles factors affecting trust and, in general, HMI effectiveness \cite{ekman2018creating}. Specific sections were included  to  test  each  of  these  factors.  The questionnaire includes both general questions, that could be re-used across different driving scenarios, as well as questions that are more specific to HMI and to the simulated scenario. We focused our attention on those aspects that are more relevant for establishing trust in an initial learning phase, where the user gets acquainted with the system. When possible, questions were mutuated from validated tools such as  Simulator Sickness Questionnaire (SSQ) \cite{kennedy1993simulator}, the Situation Awareness Rating Technique (SART) \cite{taylor2017situational} and the NASA Task Load Index (NASA-TLX) \cite{tlx}. Questions were organized in the following sections.
\subsubsection{Health status}
\label{sec:quest_health}
VR systems may induce motion sickness and other side effects: to avoid biases, health status is collected before and after the experience using the SSQ tool.
\subsubsection{System competence}
Inspired by standard questions for the evaluation of trust in human-robot interaction (HRI), this section evaluates the perceived system's \textit{competence} across the range of driving situations explored in the simulation \cite{schaefer2016measuring}.
\subsubsection{Reaction to test events}
\label{sec:quest_events}
For each test event, the user is asked to rate four statements: \textit{1) The situation was dangerous}, \textit{2) The event took me by surprise}, \textit{3) I was able to see the potential danger before it affected the vehicle's performance}, and \textit{4) The interface provided me useful information to foresee the event}.  These questions provide complementary information to the physiological signals and disentangle the effect of the specific event from the HMI.
\subsubsection{Situational awareness}
This section was inspired by the SART tool, focusing on dimensions (quality, quantity and familiarity) that pertain to comprehensibility. Here, quality refers to the usefulness with respect to clarifying system's intentions. Quality and quantity were evaluated for each element of the HMI, e.g., bounding boxes, navigation lines, etc. 
\subsubsection{Cognitive load}
This section was adapted from the NASA-TLX evaluation tool.
\subsubsection{Overall user experience}
This section investigates general aspects regarding the mental model, and is concluded with a direct question about trust. Predisposition towards participating in an AD experience was also assessed before and after the simulation. 
\subsubsection{Immersion and presence}
Immersion, presence and simulation fidelity were evaluated by adapting the relevant sections from the VRUSE questionnaire \cite{kalawsky1999vruse}, an established technique to measure usability of VR applications.

All questions were in Italian and had to be rated on a 1--5 Likert scale. Sections \textit{Reaction to test events} and \textit{Situational awareness} included snapshots of the test events and the HMI elements, respectively; the questionnaire was adapted for each test group with snapshots from the specific HUD version. The complete questionnaire (SEL version) is available at \href{https://forms.gle/CpSYZc729fho7gy86}{https://forms.gle/CpSYZc729fho7gy86}.

\section{Experiments}

\label{sec:experiments}

\subsection{Data Acquisition}

Healthy individuals (e.g. with no impairing chronic or acute illnesses at the time of the acquisition) with a valid driving license were recruited to participate in the virtual driving experiment. Participation was voluntary and no monetary compensation was provided. Study participants were randomly assigned to either the OMN or SEL HUD group.  All acquisitions were performed within one week. 

The test phase began for each subject with a brief explanation of the test session. Health status, demographic information and general disposition towards AD systems were collected before starting the simulation. Two baseline signals were also acquired: one minute at rest, and one minute after placing the VR headset. After the simulation, the final questionnaire was administered and the experience debriefed.

The GSR was recorded through an ad-hoc device based on the Groove GSR Sensor \cite{groove} and a Raspberry Pi 3 board. The acquisition module was implemented in Python. An external Analog to Digital Converter (MCP3008 \cite{mcp}) was used to connect the output of the sensor to the board via the Raspberry's Serial Peripheral Interface (SPI). The sampling frequency was set to 256 Hz in order to separate the two components of the GSR signal \cite{figner2011using}. Due to inter-subject variability, the GSR may saturate during the analog to digital conversion: therefore, during the initial baseline acquisition, the converter was manually calibrated by adjusting the resistor until the output fell in the 200--512 a.u. range. The sensors were applied on the fingers of the non-dominant hand, after washing the hands. Postprocessing and feature extraction was implemented in Python 3.6.5 and the SciPy library for filtering; all calculations were performed on an HP Pavilion, Intel Core i5-3230M CPU. 

\subsection{Statistical Analysis}

A two-way factorial Analysis of Variance (ANOVA) was conducted to examine the main effect of HUD as well as the interaction effect between event and HUD type on each GSR feature. A mixed design was employed with the HUD type as the between-groups factor and the event as the within-subjects factor. Post-hoc comparison between the different events and HUD types was performed applying Bonferroni correction.

Questionnaire data was analyzed separately for each group of questions. Event-related questions were analyzed using a two-way factorial ANOVA, using the same design of the GSR feature. Outcomes of the other questions were compared between the OMN and SEL  groups using the Mann-Whitney U-test for categorical data. A $p$-value of .05 or lower was considered to indicate a statistically significant difference. Statistical analysis was performed using SPSS v20, whereas signal analysis and feature extraction were coded in Python. 

\subsection{Participants' Characteristics}
Thirty-nine subjects volunteered to participate in the study. One subject with excessive motion sickness was excluded from the data-set, as symptoms would bias the physiological response \cite{patrao2015deal}. A total of 38 subjects (25 male, 13 female, mean age 23.9) were included in the  analysis. GSR data was not available for 8 subjects due to failures in the recording equipment.  Most of the subjects reported using VR or driving simulators ``never'' or ``rarely'' (30/38 and 34/38, respectively).

\subsection{Quantitative Measurements}
\begin{figure}[th]
    \centering
    \subfloat[OMN]{\includegraphics[height=3.4cm]{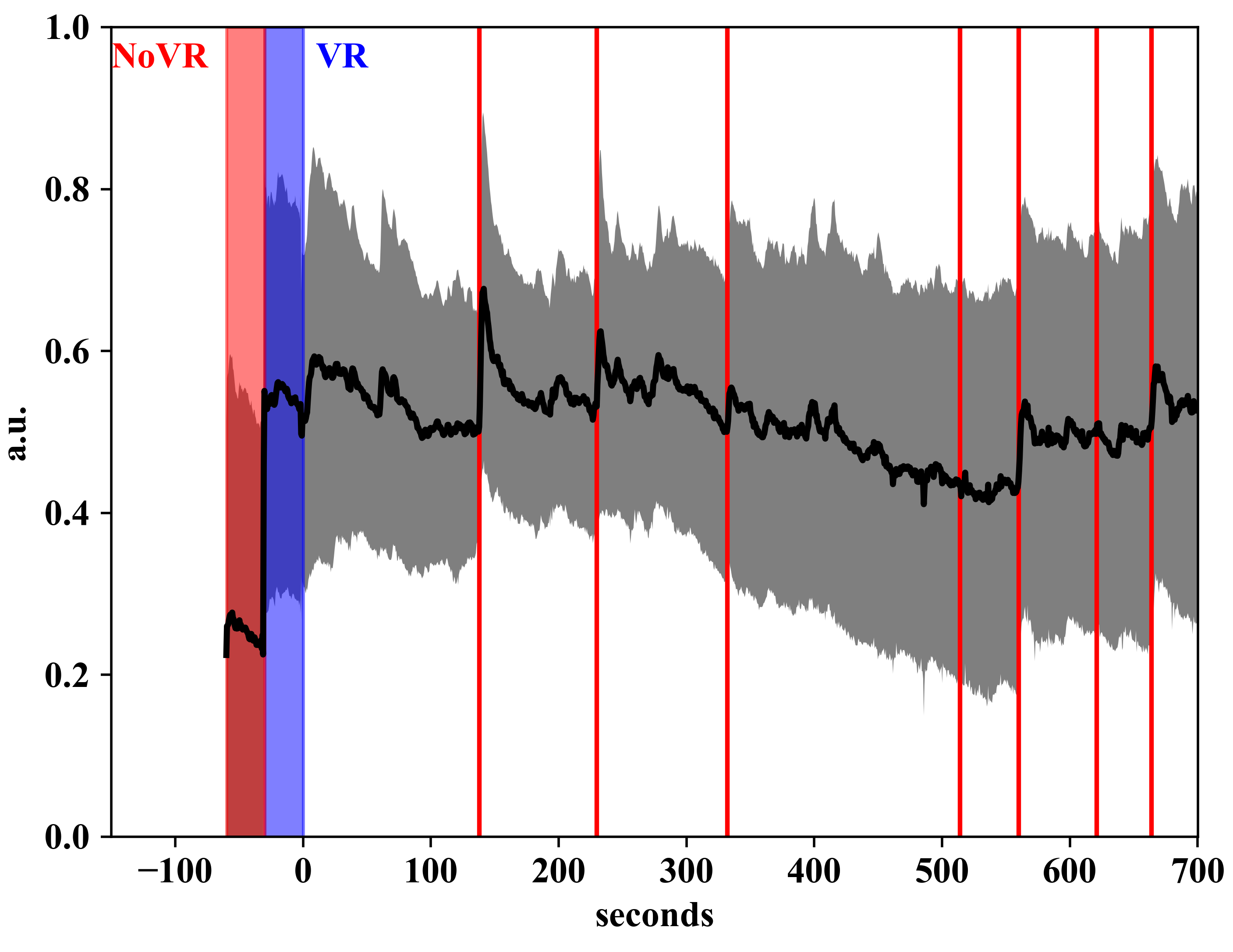}}
    \subfloat[SEL]{\includegraphics[height=3.4cm]{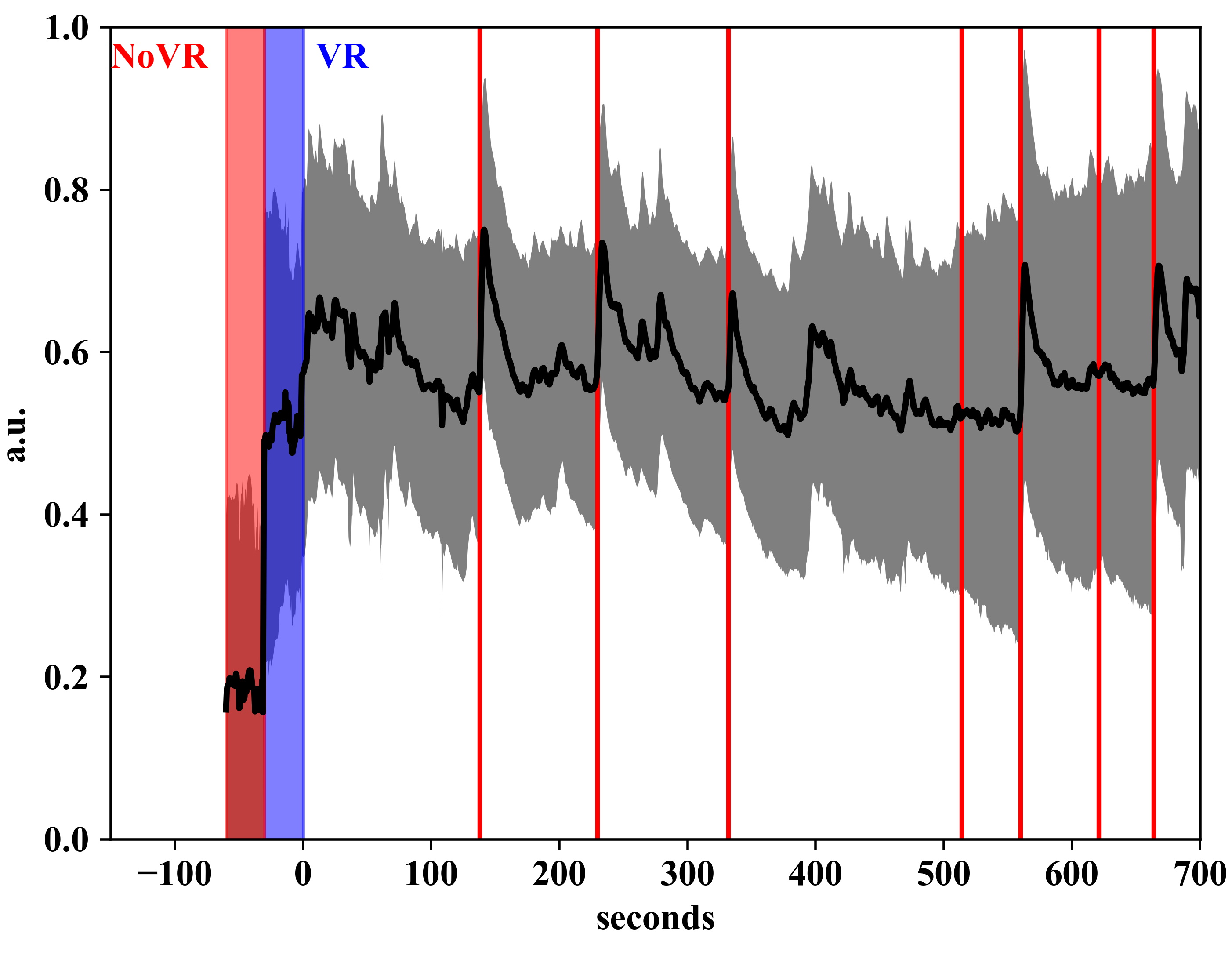}}
    \\
     \subfloat[OMN]{\includegraphics[width=0.5\columnwidth]{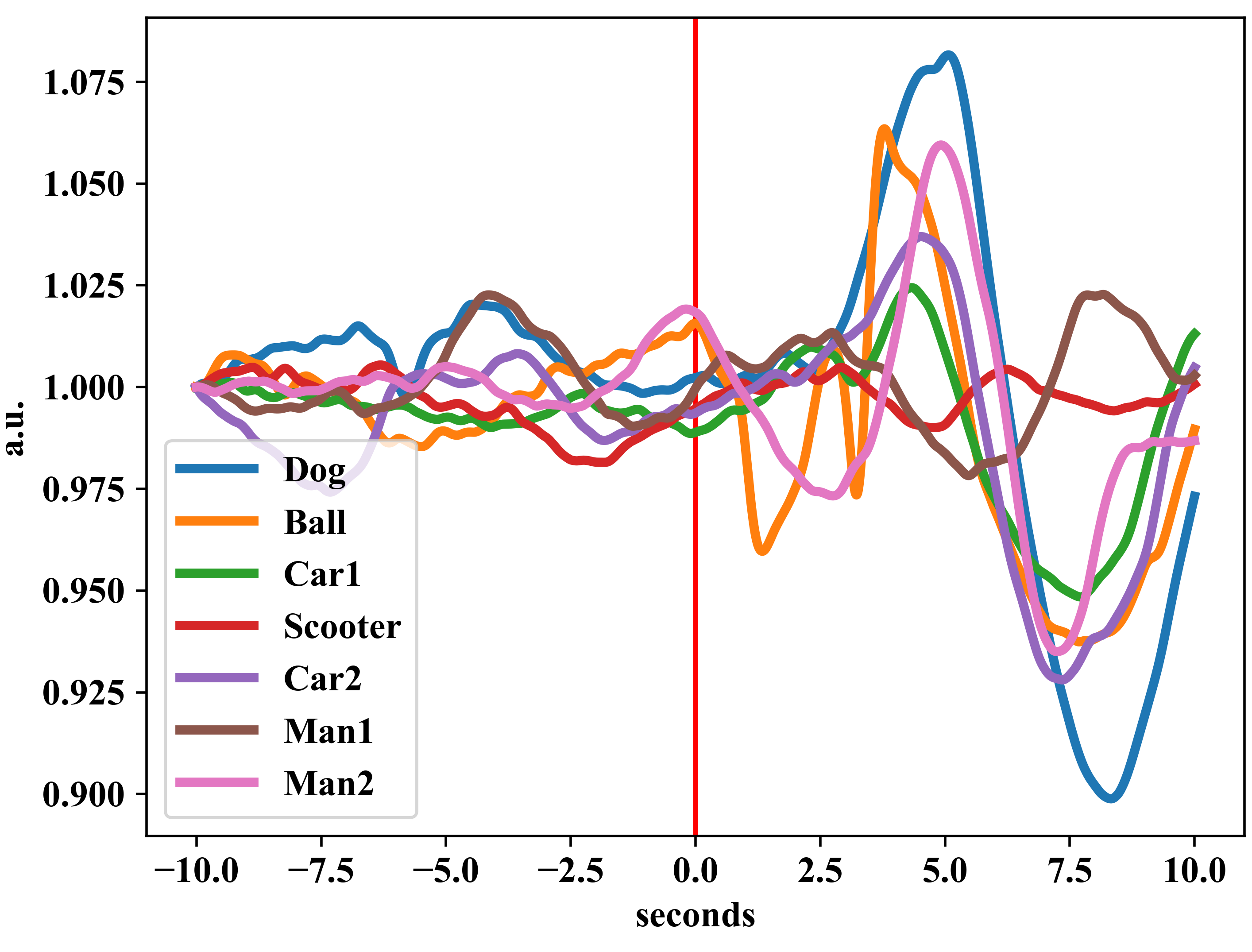}}
    \subfloat[SEL]{\includegraphics[width=0.5\columnwidth]{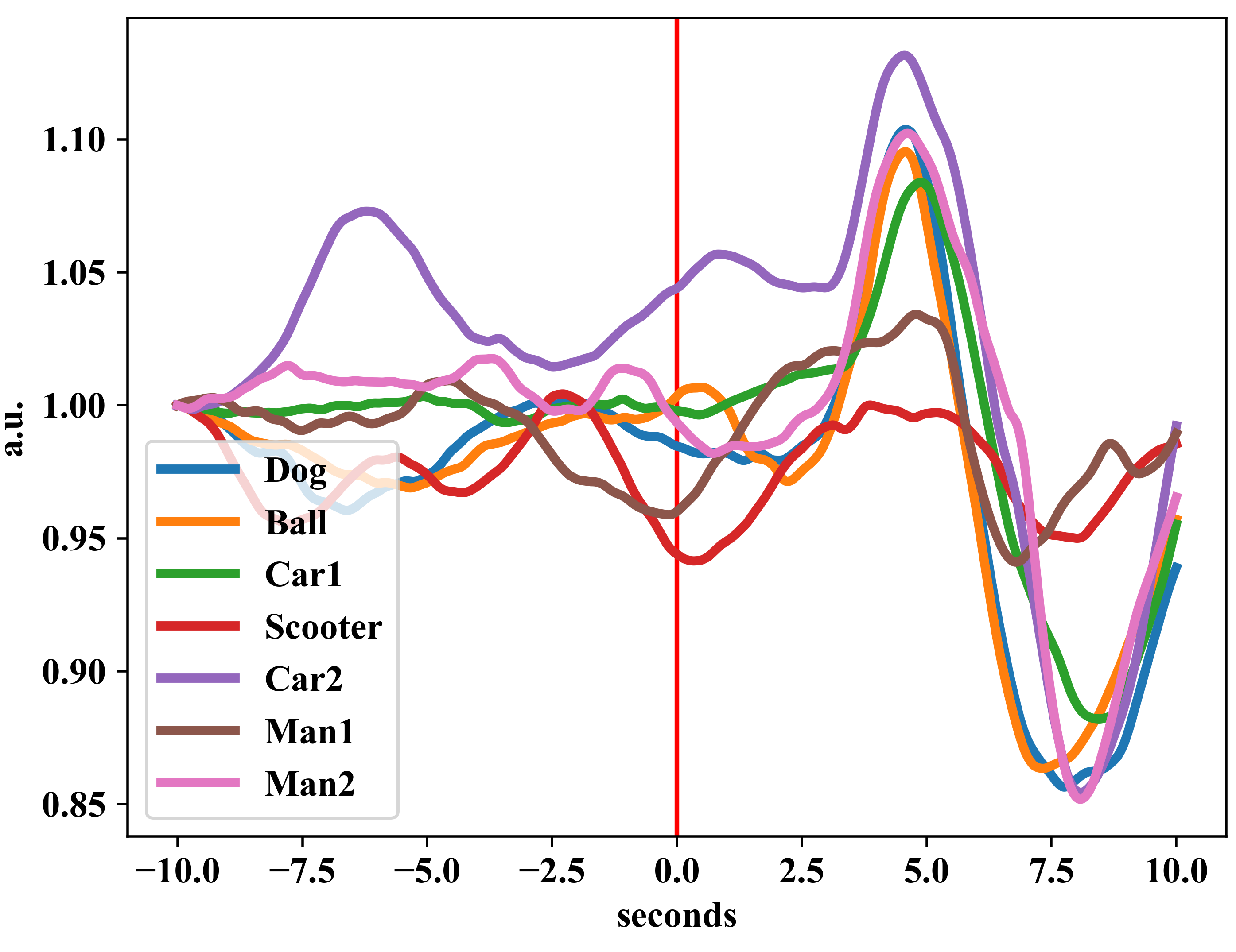}}

 \caption{Normalized raw GSR signal for OMN (a) and SEL (b) interfaces; baseline is collected prior to the experience and red lines represent test events in the simulation. Average SCR curves over all subjects in the 10s before and after the event for all the test events for OMN (c) and SEL (d) interfaces.}
    \label{fig:scr_event}
\end{figure}

The normalized GSR signals averaged over all study subjects within each group are reported in Fig. \ref{fig:scr_event}(a)--(b). All subjects showed an increase in baseline GSR in VR. Moreover, a noticeable peak in the GSR occurred for most events in the test scenario. Fig. \ref{fig:scr_event}(c)--(d) show the mean SCR curve for each event. Each curve is extracted for a time window of \textpm 10s centered at each event; within each window, all samples are divided by the first value to highlight changes.

From the SCR and GSR curves, different features have been extracted, as defined in Section \ref{sec:feature}. We here report in detail the two-way ANOVA results for the $\Delta P2P$ feature. The main effect of HUD was significant, F(1,28)=4.72, $p$=.039, indicating a statistically significant difference between the OMN and SEL interfaces. The main effect of event was also statistically significant, F(6,168)=13.9, $p$\textless.001. We did not find a significant interaction  between HUD and event, F(6,168)=1.74, $p$=.115; hence, post-hoc analyses were conducted on each main effect separately. 

\begin{figure}[t]
    \centering
    \includegraphics[width=0.80\columnwidth]{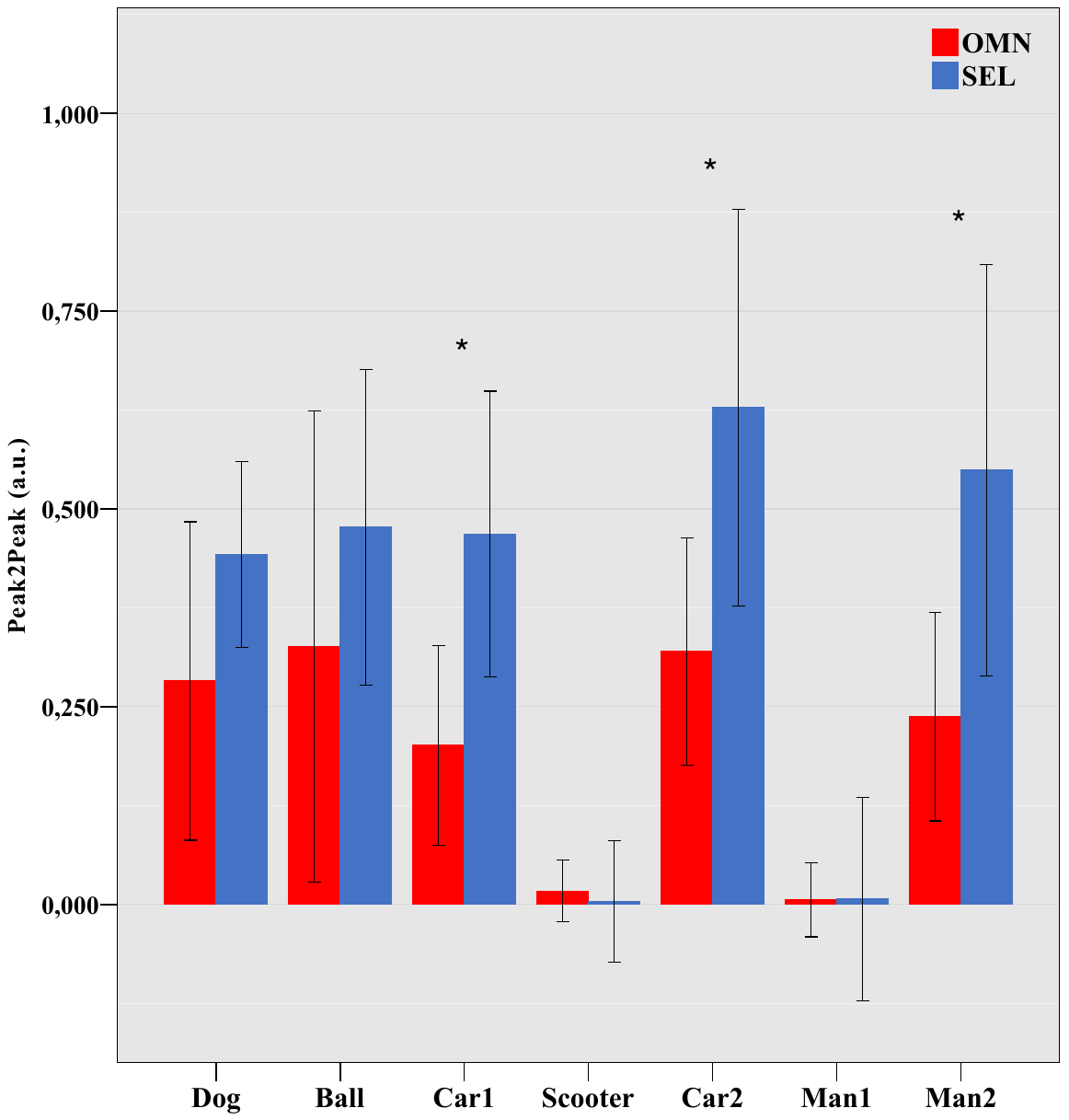}
    \caption{$\Delta P2P$ feature, for all events, with OMN and SEL. $^*$ $p$-value\textless .05.}
    \label{fig:scr_comparison}
\end{figure}

The mean and standard error of the SCR feature for each event and for each HUD are reported in Fig. \ref{fig:scr_comparison}. At post-hoc analysis, the SEL HUD consistently showed higher emotional arousal for Car1 ($p$=.022), Car2 ($p$=.042) and Man2 ($p$=.041) events. For the first two events in the timeline, a positive trend could be observed ($p$=.181 and $p$=.409). For the Scooter and Man1 events, which elicit no emotional arousal, differences were not statistically significant ($p$=.759 and $p$=.990).

All GSR features showed a significant main effect of HUD: $\Delta Max$, F(1,28)=8.53, $p$=.007;  $\Delta GSR_{Mean}$, F(1,28)=9.36 $p$=.005, and $GSR_{Acc}$, F(1,28)=9.02, $p$=.006. Likewise, a significant main effect of event was always found ($p < .001$), with no significant interaction between HUD and event. At post-hoc analysis, results for the $\Delta Max$ feature were comparable to $\Delta P2P$, whereas $\Delta GSR_{Mean}$ and $\Delta GSR_{Acc}$ features reported significant differences ($p$<.05) for Ball, Car1 and Car2 events, instead of Car, Car2 and Man2 events.

For each HUD and each test event, $\Delta P2P$ \textit{Pre} (10s before the event) and \textit{Post} (10s after the event) values were tested for differences using two-tailed t-tests. With the SEL HUD, all events showed a significant increase in SCR ($p$<.001), except for Scooter ($p$=.927) and Man1 ($p$=.920). Very similar results were obtained with the OMN HUD: the Dog ($p$=.014), Ball ($p$=.046), Car1 ($p$=.007), Car2 ($p$=.001) and Man2 ($p$=.003) events showed a significant effect on SCR, whereas Scooter ($p$=.142) and Man1 ($p$=.422) did not. Results for other GSR features, here omitted for brevity, were also consistent.

\subsection{Questionnaire Results}
\label{sec:questionnaire_results}

Only one subject was excluded from this analysis due to high motion sickness; the other subjects did not report excessive symptoms (nausea rating M=1.26, SD=.54).

\begin{figure}[t]
    \centering
    \includegraphics[width=0.90\columnwidth]{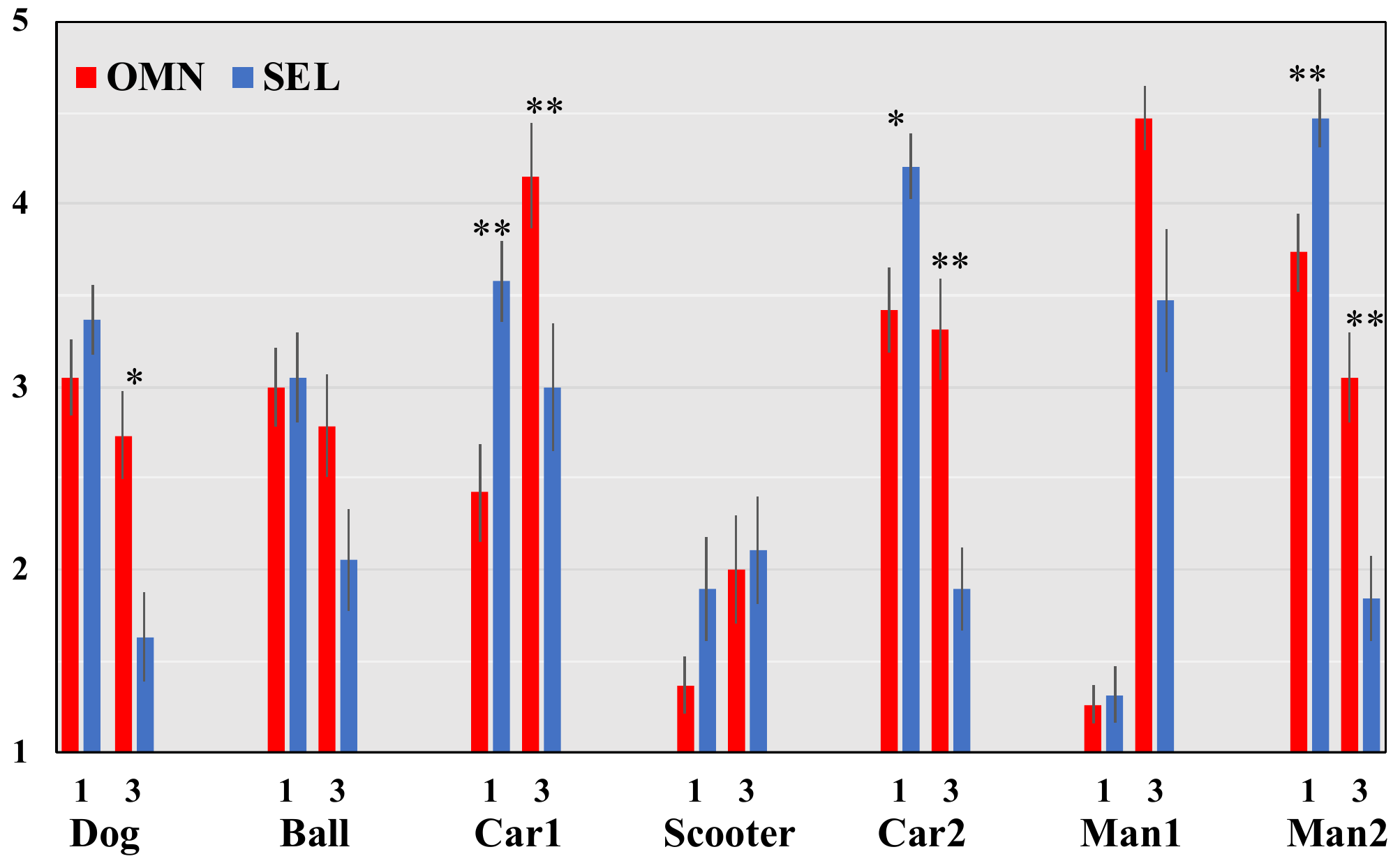}
    \caption{Subjective measurements for questionnaire section \textit{Reaction to test events}. Label 1 refers to the question which evaluates the risk perception, on a scale from 1 (low risk) to 5 (high risk); label 3 refers to the question that evaluates if and how the individual previously noticed the dangerous situation, on a scale from 1 (not previously noticed) to 5 (previously noticed). Each test event is considered separately. $^*$$p$-value \textless.05, $^{**}$$p$-value \textless .01.}
    \label{fig:quest1}
\end{figure}

Subjective ratings for test events are reported in Fig. \ref{fig:quest1}. Four statements were included for each test event, as detailed in Section \ref{sec:quest_events}; for the sake of clarity, only question 1 (which evaluates the risk) and question 3 (which evaluates the ability to detect the potential danger in advance) are included in the plots, as answers to questions 2 and 4 were very similar. At two-way ANOVA, the main effect of both HUD, F(1,36)=15.91, $p$<.001, and event, F(6,216)=54.05, $p$<.001, on the perceived risk (question 1) were statistically significant. Interaction between the two factors failed to reach statistical significance, F(6,216)=2.05, $p$=.060. Regarding the ability to identify dangerous situations in advance (question 3), the main effect of both HUD, F(1,36)=28.08, $p$<.001, and event, F(6,216)=14.78, $p$<.001, were statistically significant, without a significant interaction, F(6,216)=1.75, $p=.112$. 

Since events are the same in both groups, we attribute the difference in perceived risk to the greater ability of the OMN interface to convey information about the vehicle's surroundings before critical situations occur. At post-hoc analysis, differences were statistically significant for Car1 ($p$=.003), Car2 ($p$=.017) and Man2 ($p$=.008) events, and a positive trend was observed for Dog ($p$=.134) and Ball ($p$=.872) events. 

\begin{figure}
    \centering
 \subfloat[]{\includegraphics[height=3.3cm]{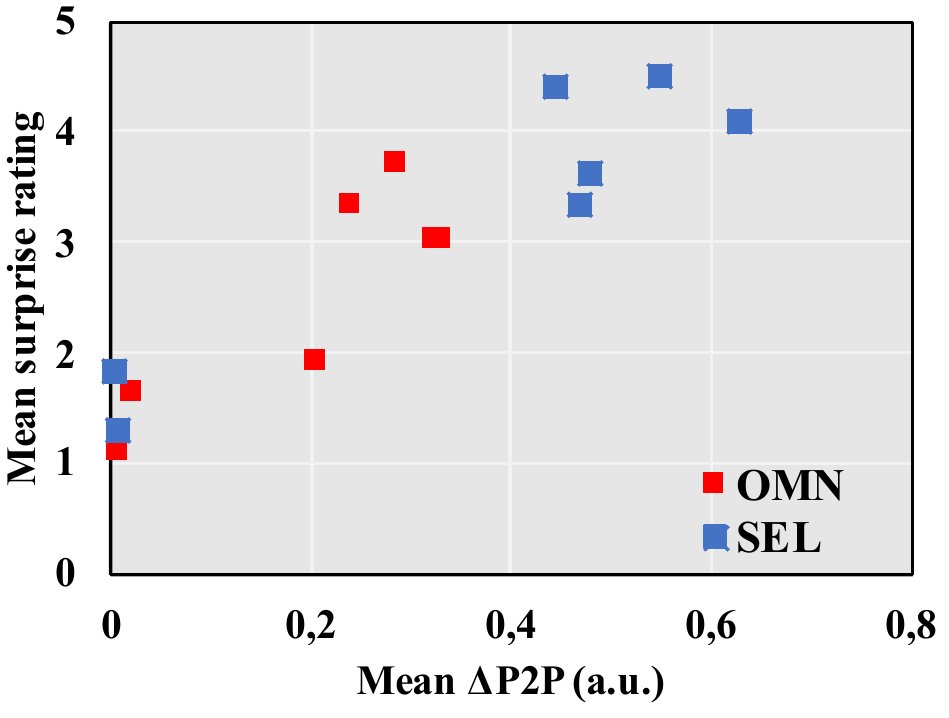}}
    \subfloat[]{\includegraphics[height=3.3cm]{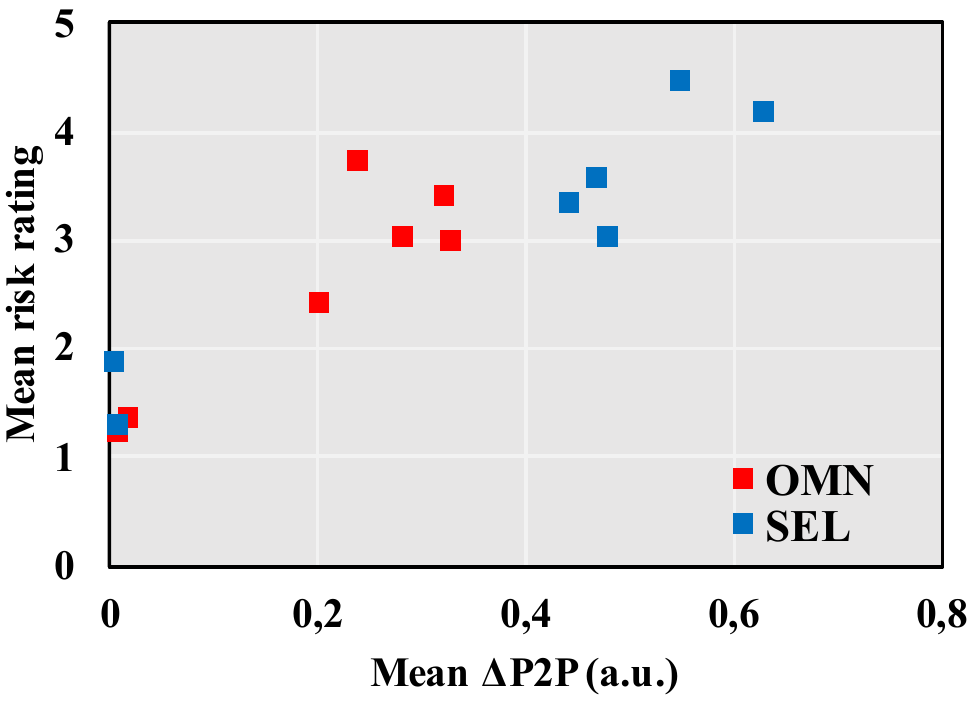}}
    \caption{Comparison of subjective vs. objective ratings. For subjective measurements, the
     average perceived risk (a) and the average surprise rating (b) are reported (where the latter refers to the extent to which the user was taken by surprise by the event). The mean $\Delta P2P$ feature is reported as objective rating. Each data point corresponds to a specific test event. }
    \label{fig:quantvsqual}
\end{figure}

For each event, questionnaire ratings and GSR features values were compared by using multiple linear regression; by attempting to predict the average GSR outcome ($\Delta P2P$) from the average questionnaire ratings, we can desume the degree of similarity between the two measurements. A per-subject analysis was not attempted, given the limited sample size. A statistically significant regression equation was found, F(4,9)=14.34, $p$=.0007, with an adjusted $R^2$ of 0.804, which indicates that roughly 50\% of the variance of the GSR can be explained by the questionnaires. Individual factors failed to reach statistical significance, but the strongest trends were observed for the perceived level of risk (coefficient 0.111, $p$=.29) and the element of surprise (coefficient 0.112, $p$=.29), which are presented in the scatter plots of Fig. \ref{fig:quantvsqual}.

Subjects generally found the vehicle's driving skills adequate (SEL M=4.53, SD=0.61, OMN (M=4.68, SD=0.48, $p$=.556). In the SEL group, subjects reported more often that the vehicle faced difficulties with unexpected changes in the environment (SEL M=1.68, SD=0.75 and OMN M=1.21, SD=0.42, $p$=.41); such differences can only be attributed to the HUD, considering that the vehicle's behavior was exactly the same in both experiences. 

Displaying more information may result in an excessive cognitive load. Indeed, subjects in the OMN group more often rated the amount of information provided by the interface as excessive (OMN M=2.1, SD=0.229, SEL M=1.05, SD=0.809, $p$<.001), whereas comprehensibility was rated adequate for both the interfaces ($p$=.908). On average, the UX was satisfactory for both the interfaces, and the information provided by the HUD was considered useful (SEL M=4.16, SD=0.69, OMN M=4.84, SD=0.38, $p$=.001). Participants in the OMN group reported that the information was more useful in order to understand why the vehicle made a decision (SEL M=4.26, SD=1.05, OMN M=4.84, SD=0.38, $p$=.055) and to feel in general at ease (SEL M=3.79, SD=1.08, OMN M=4.68, SD=.59, $p$=.003), as well as that the vehicle seemed to have greater control on the external environment (SEL M=3.79, SD=1.08, OMN M=4.84, SD=0.38, p$<.001$). Overall, the OMN HUD was more helpful in anticipating potential dangers (SEL M=2.42, SD=0.61, OMN M=4.10, SD=0.57, $p$<.001). Subjects reported a high sense of immersion (M=4.50, SD=0.73) and presence (M=4.37, SD=0.59), with no significant difference between the two groups. 

\begin{figure}
    \centering
    \includegraphics[width=0.8\columnwidth]{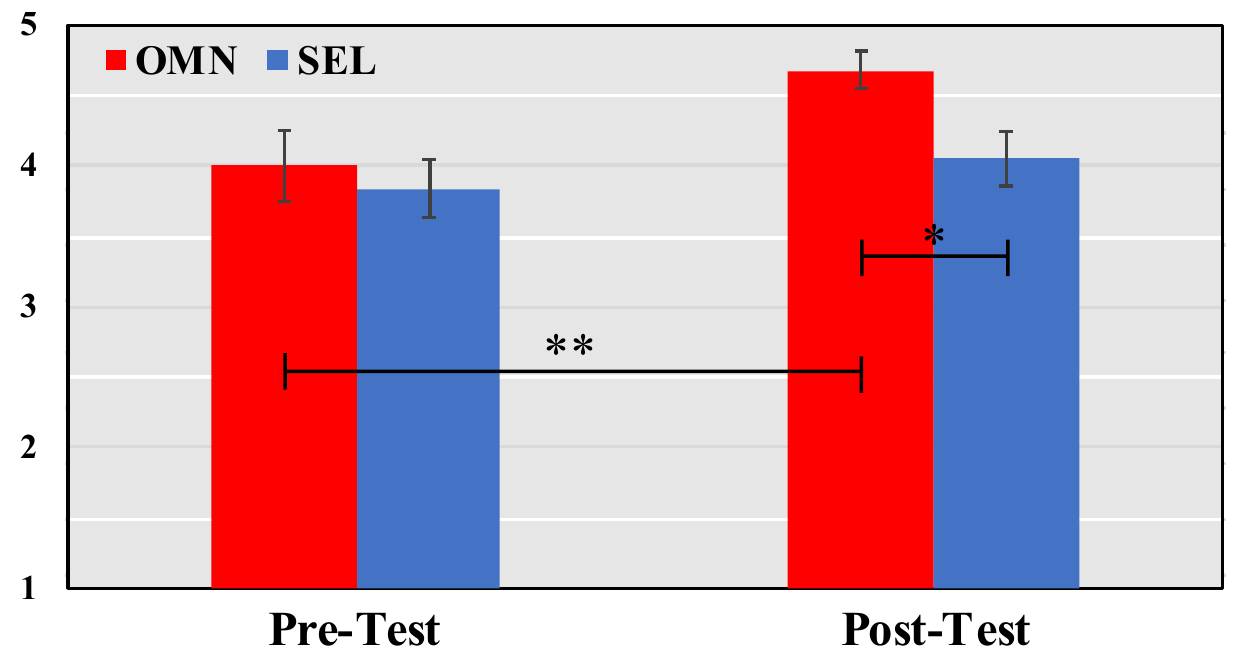}
    \caption{Disposition towards participating in a real AD experience. On the left, the pre-test answer for the SEL and OMN interfaces; on the right, the post-test answer. Scale from 1 (absolutely negative) to 5 (absolutely positive). Mann-Whitney U-tests between pre- and post-test answers are shown, $^*$$p$-value\textless.05, $^{**}$$p$-value \textless0.01.}
    \label{fig:pre_posthud}
\end{figure}

Finally, users in the OMN group were better disposed towards participating in a real AD experience (OMN M=4.68, SD=0.58, SEL M=4.05, SD=0.85, $p$=.012). As shown in Fig. \ref{fig:pre_posthud}, prior to the experiment all participants were mildly optimistic, but after experiencing the OMN HUD, attitude towards the technology markedly improved (M=4.0 vs. M=4.68, $p$=.002). 

Complete  data is provided in the Supplemental material.

\section{Discussion}
We here  proposed a methodology to validate the UX in AD systems based on continuous, quantitative information gathered from physiological signals while the user is immersed in a VR driving simulation. Our methodology is exemplified by the comparison of two AR-HUD-based interfaces which differ in the amount of information displayed to the users. 

By controlling all aspects of the simulated environment, we were able to disentangle the effect of very specific design choices and measure their impact on the overall UX. It must be stressed that the only difference between the two groups was the information displayed by the HUD, as the simulation was otherwise identical; study groups were also  homogeneous in terms of age, sex and ethnicity.  

Our results confirmed that providing ``why'' information is important to reassure the user of the system's competence and to promote trust and situational awareness \cite{koo2015did, ekman2018creating}. To the best of our knowledge, ours is the first contribution to evaluate a realistic HUD displaying a wide range of visual and auditory cues about the vehicle and its surroundings, as it is expected in future AVs. Given the number of objects involved in realistic scenarios, an omni-comprehensive (OMN) display could lead to an excessive cognitive load. A possible way to reduce information load, which we denoted as selective (SEL), is to display only the most relevant visual cues in the current context.  Indeed, our results indicated that the users found information displayed by the OMN HUD slightly excessive, although acceptable in both cases, but this was compensated by a less stressful driving experience, as confirmed both by subjective and objective measures. 

This difference is especially evident when potentially dangerous events occur, such as a pedestrian crossing the street at the last minute. It is worth noting how the HMI influenced the perception of external events, on one hand, and of the vehicle's performance, on the other hand, despite the fact that the simulated scenario was identical in those respects. For instance, users in the OMN group perceived the vehicle as better equipped to deal with unexpected changes in the environment. We argue that this difference arose as a consequence of the mental model that users formed: as the information provided by the HUD allowed users to better anticipate dangerous situations, they projected this feeling onto the AV as well. 

Our results have important implications for AI research in AD, and specifically for the sensory sub-systems, as HMI constraints need to be considered in their design. For instance, end-to-end training from sensory input to planning does not explicitly extract all the information that was included in this simulated HMI \cite{bojarski2016end}.  In our simulation, information displayed by the SEL HUD was chosen based on a set of heuristics that could be further improved by exploiting a more advanced AI, such as the ability to predict the motion of objects and pedestrians to foresee potentially dangerous situations before they actually affect the vehicle's trajectory. 

In this study, we have sought to be as independent as possible from specific AD systems, e.g., by simulating perfect vehicle sensing capabilities. Our conclusions are thus unaffected by potential errors or misses in the AD object detection system. The proposed methodology could certainly be employed to test other types of autonomous vehicles and their underlying AI systems, by changing the modeled interior and/or behavior. It would also be possible to investigate how possible errors may affect the UX and trust.

The proposed scenario is certainly representative of the learning phase as defined in \cite{ekman2018creating}. Information display by the HUD is particularly relevant in this initial phase, when the user is still forming a mental model of how the AD system works. Our results may not apply entirely to the performance phase, in which the user has observed the AD system for a prolonged period of time. However, the unexpected events or accidents which we simulate, while rare, can have a profound effect on trust, both at the individual and collective level. It should be noticed that trust begins to form even before the first interaction with the system, e.g. based on information from the media, or personal preferences \cite{LeeSee, ekman2018creating}. This was evident in our study where, initially, many subjects were not willing to participate in a real AD experience. However, participating in the VR experience, and being exposed to an informative interface, significantly improved their acceptance towards AD systems. In a simulated setting, all AD technologies, as well as all types of events, can be recreated, opening interesting opportunities for ``training'' future users of AV technology. 

GSR proved capable of detecting user's stress in response to potentially dangerous events, in line with previous literature results which, however, were obtained in the context of manual or partially automated driving \cite{healey2005detecting, eudave2017physiological}. Notably, differences in HMI design were reflected in observable changes in GSR levels, even when using consumer electronics sensors. The GSR response was correlated to the perceived risk as measured by subjective questionnaires, as well as to the ``surprise'' factor, which depends on the HMI.  We here focused on the response to specific events, but the methodology could be extended to extract features that characterize the entire experience \cite{healey2005detecting}.

\section{Conclusions and Future Work}
In this work we proposed a methodology to validate the UX in AD systems based on continuous, quantitative information gathered from physiological signals while the user is immersed in a VR driving simulation. Its effectiveness was shown in the context of HMI design, and specifically applied to the comparison of HUD-based interfaces for AVs that provides visual cues about the vehicle's sensory and planning systems. We explored in this exemplification the role of HMI in eliciting a sense of trust and safeness in AD systems, as this will be key for humans to relinquish control of the vehicle.

The proposed methodology relies on physiological signals (GSR in this specific embodiment) to provide a continuous, quantitative and objective feedback. This is particularly relevant for simulation of AD systems, as objective measures in driving research are traditionally based on driver's performance and behavior. A limitation of GSR is that it measures arousal, but is a poor indicator of valence. In our specific case, the experience was engineered to elicit a sense of distress and, hence, a positive valence was excluded. In the future, this lack could be overcome by including other sensors, e.g., to measure the HR,  other types of features that reflect different characteristics of the UX, as well as machine learning models to more accurately detect the passengers' affective state. 

It should be noticed that the increasing adoption of wearable devices like smart watches incorporating a growing set of health sensors will open additional opportunities for AVs' personalization; anthropomorphism, customization and adaptivity are also important factors for trust-worthy HMI \cite{ekman2018creating}. While the physiological response (1--5s in the case of GSR) is too slow to be exploited for actual driving, it could be used to customize various aspects of the HMI, like the quantity and quality of information displayed, and of the overall driving experience. 

The proposed methodology for testing could be extended to cover also the above scenarios as well as other aspects of the UX (e.g., considering not just in-vehicle scenarios, but also vehicle-to-pedestrian interactions \cite{pedestrianvr}, long-term performance \cite{ekman2018creating}, etc.), by adjusting the simulation, the HMI and/or the vehicle's AI as needed.

\section*{Acknowledgment}
The authors want to thank Dario Doronzo and Antonello Laurino for their contributions on the system implementation. This research was partly supported by the VR@Polito lab.

\ifCLASSOPTIONcaptionsoff
  \newpage
\fi



\bibliographystyle{IEEEtran}
\bibliography{bibtex/bib/IEEEabrv.bib,bibtex/bib/HMI_Vehicular_bib}
%
%

\vskip -2\baselineskip plus -1fil
\begin{IEEEbiography}[{\includegraphics[width=1in,height=1.25in,clip,keepaspectratio]{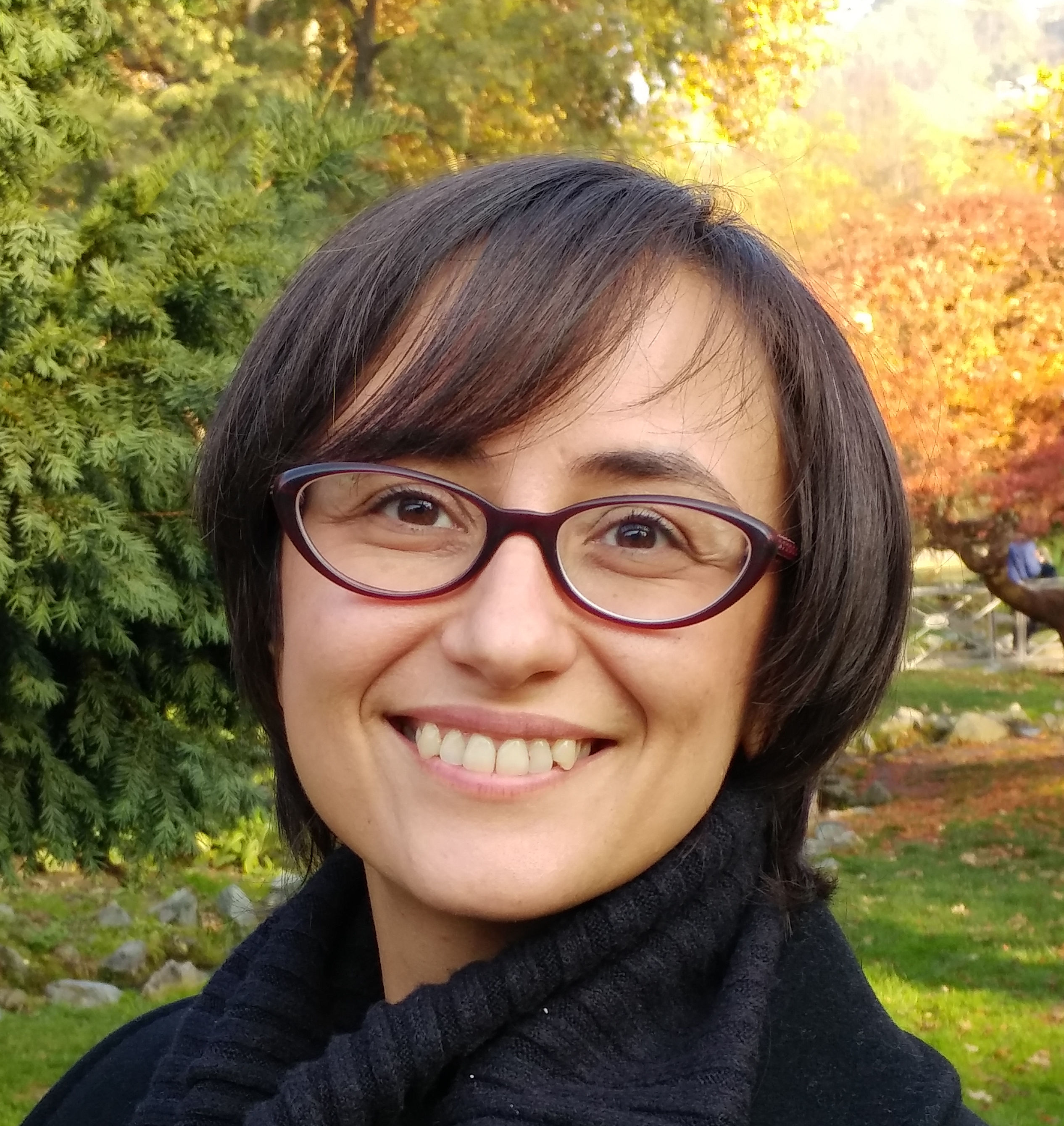}}]{Lia Morra}
 received the M.Sc. and the Ph.D. degrees in computer engineering from Politecnico di Torino, Italy, in 2002 and 2006. Currently, she is senior post-doctoral fellow at the Dip. di Automatica e Informatica of Politecnico di Torino. Her research interests include computer vision, pattern recognition, and machine learning. 
 \end{IEEEbiography}

\vskip -2\baselineskip  plus -1fil

\begin{IEEEbiography}[{\includegraphics[width=1in,height=1.25in,clip,keepaspectratio]{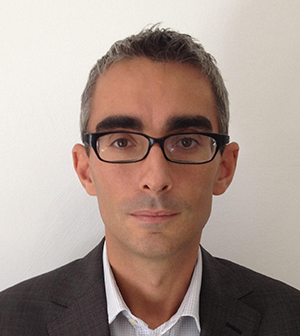}}]{Fabrizio Lamberti}
  is an associate professor at the Dip. di Automatica e Informatica of Politecnico di Torino. His research interests are manly in the areas of computer graphics, HMI and intelligent computing. He is serving as Associate Editor for the IEEE Transactions on Computers, the IEEE Transactions on Emerging Topics in Computing, the IEEE Transactions on Learning Technologies and the IEEE Transactions on Consumer Electronics. He is a Senior Member of the IEEE.
\end{IEEEbiography}

\vskip -2\baselineskip  plus -1fil

\begin{IEEEbiography}[{\includegraphics[width=1in,height=1.25in,clip,keepaspectratio]{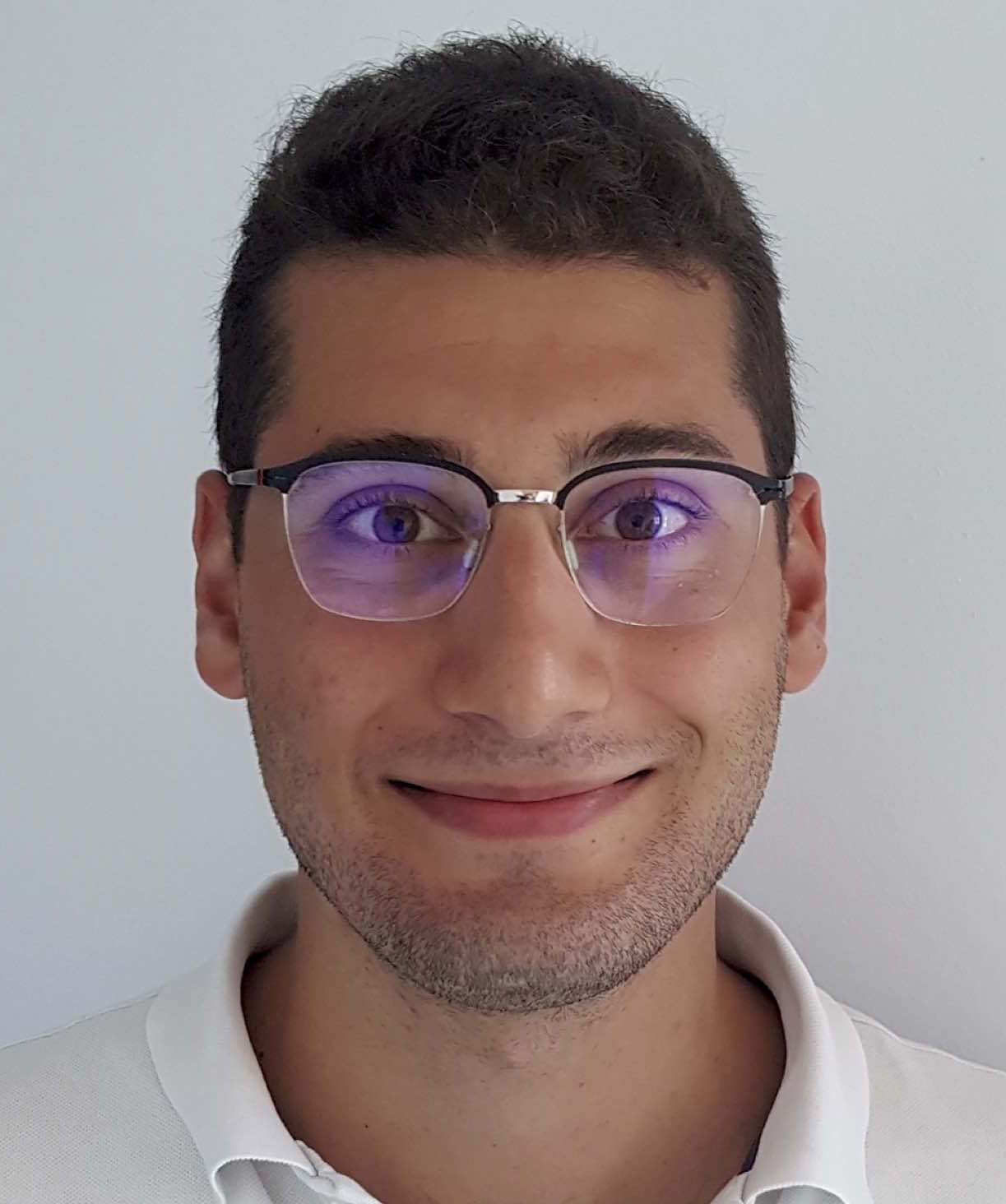}}]{F.~Gabriele~Prattic\'o}
received his M.Sc. degrees in computer engineering from Politecnico di Torino, Italy, in 2017. Currently, he is a Ph.D. student at Politecnico di Torino, where he carries out research in the areas of mixed reality, HMI, serious games and user experience design.
\end{IEEEbiography}

\vskip -2\baselineskip plus -1fil

\begin{IEEEbiography}[{\includegraphics[width=1in,height=1.25in,clip,keepaspectratio]{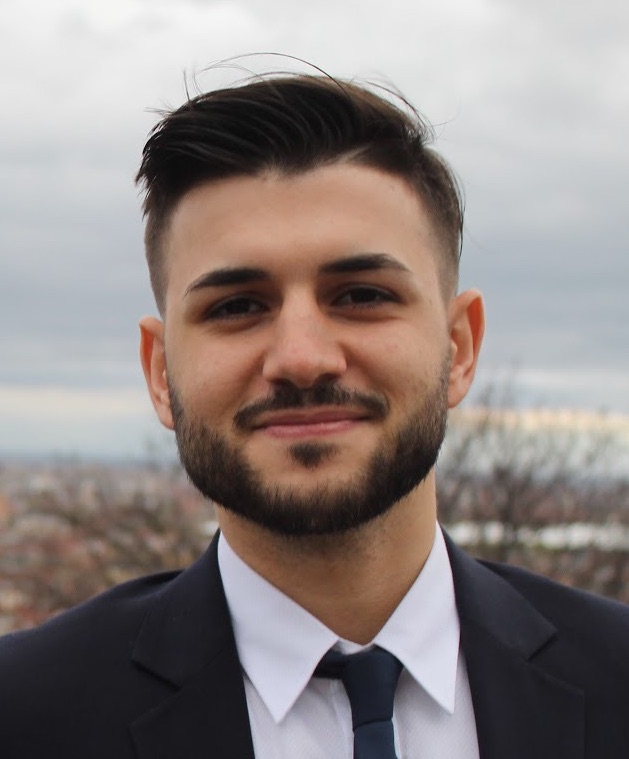}}]{Salvatore La Rosa}
 received the M.Sc. degree in biomedical engineering from Politecnico di Torino, Italy, in 2019. His major interests regard biosignal analysis, pattern recognition, machine learning and embedded systems.
\end{IEEEbiography}

\vskip -2\baselineskip  plus -1fil
\begin{IEEEbiography}[{\includegraphics[width=1in,height=1.25in,clip,keepaspectratio]{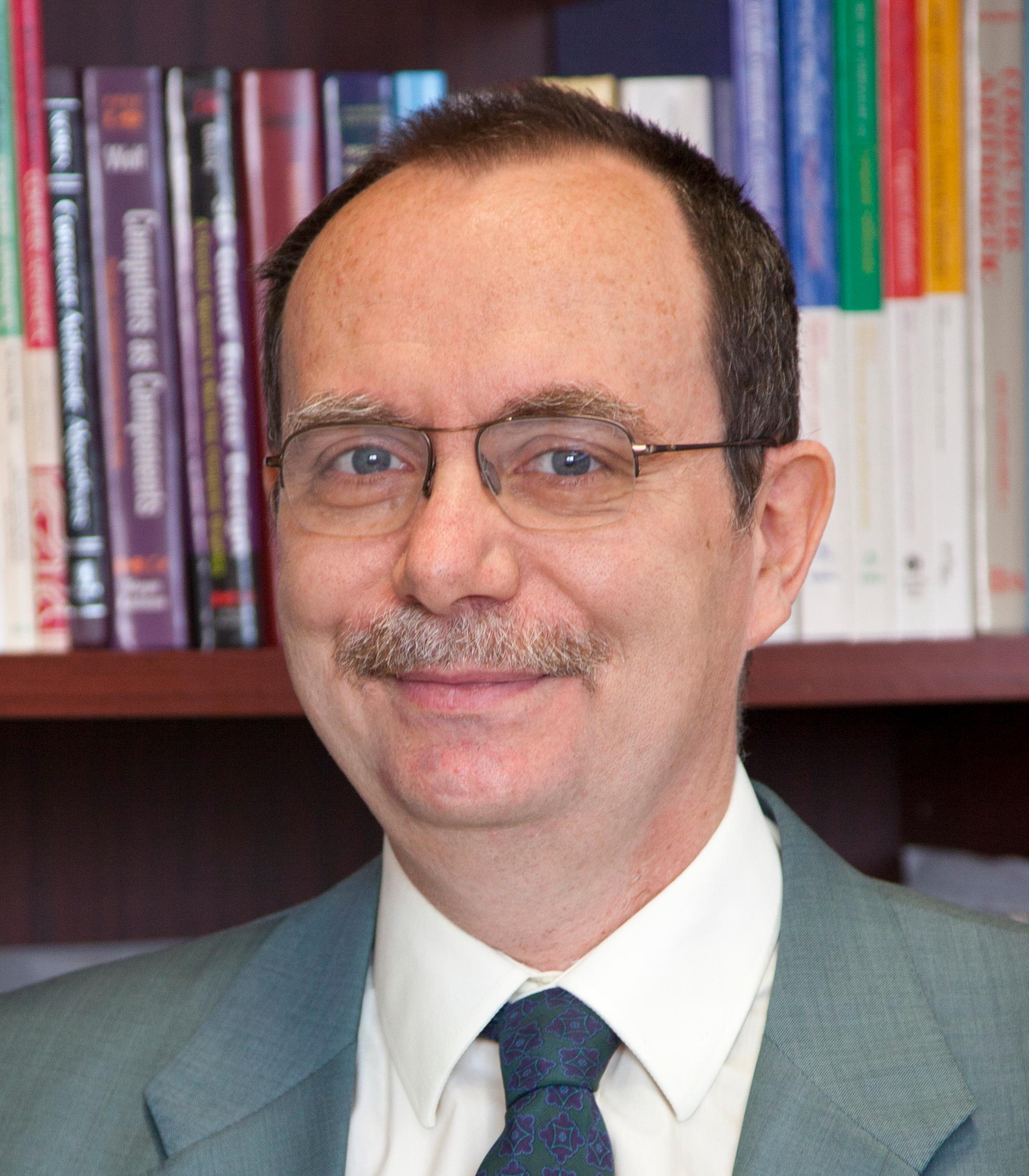}}]{Paolo Montuschi} 
is a full professor at the Dip. di Automatica e Informatica and a Member of the Board of Governors of Politecnico di Torino, Italy. His research interests include computer arithmetic, computer graphics, and intelligent systems. He is serving as 2019 Acting (interim) Editor-in-Chief of the IEEE Transactions on Emerging Topics in Computing and as the 2017-19 IEEE Computer Society Awards Chair. He is an IEEE Fellow, and a life member of the International Academy of Sciences of Turin and of IEEE Eta Kappa Nu.

\end{IEEEbiography}





\clearpage

\section*{{A}ppendix}

\begin{figure}[ht]

	\centering
	
	\subfloat[OMN]{\includegraphics[width=0.45\columnwidth]{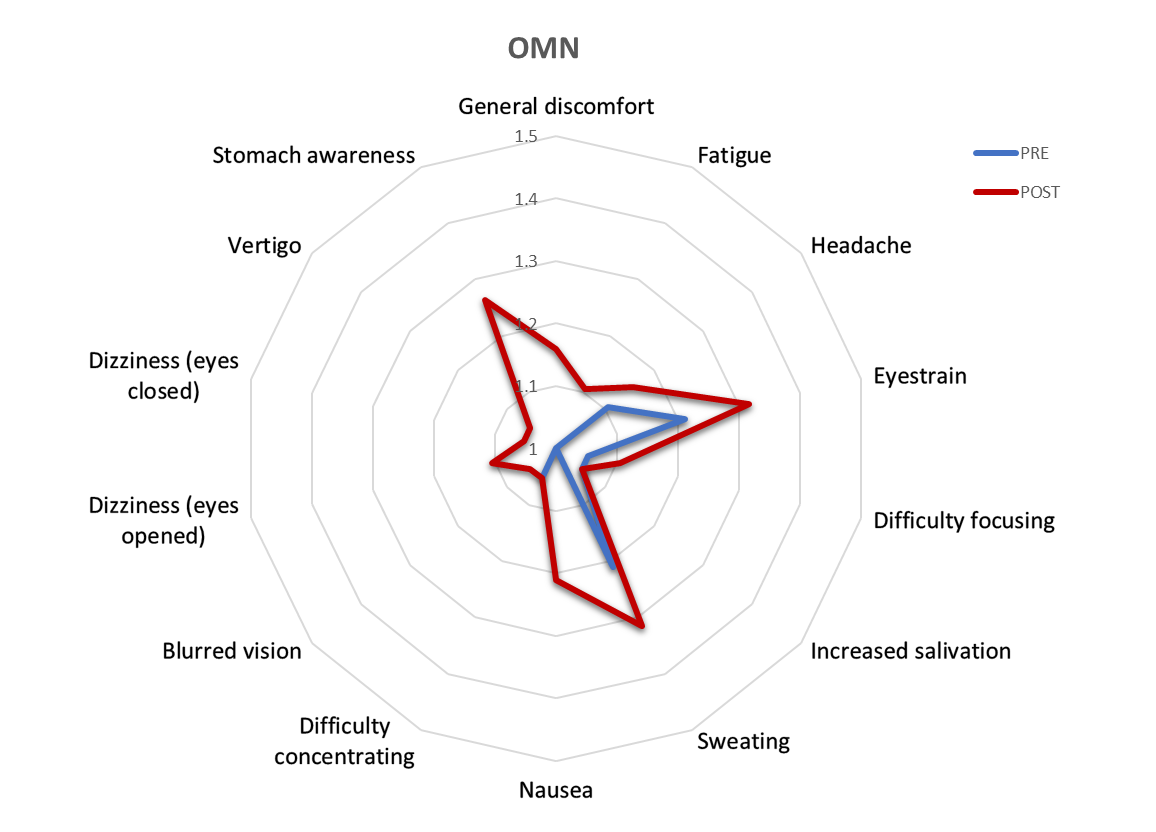}}
    \subfloat[SEL]{\includegraphics[width=0.45\columnwidth]{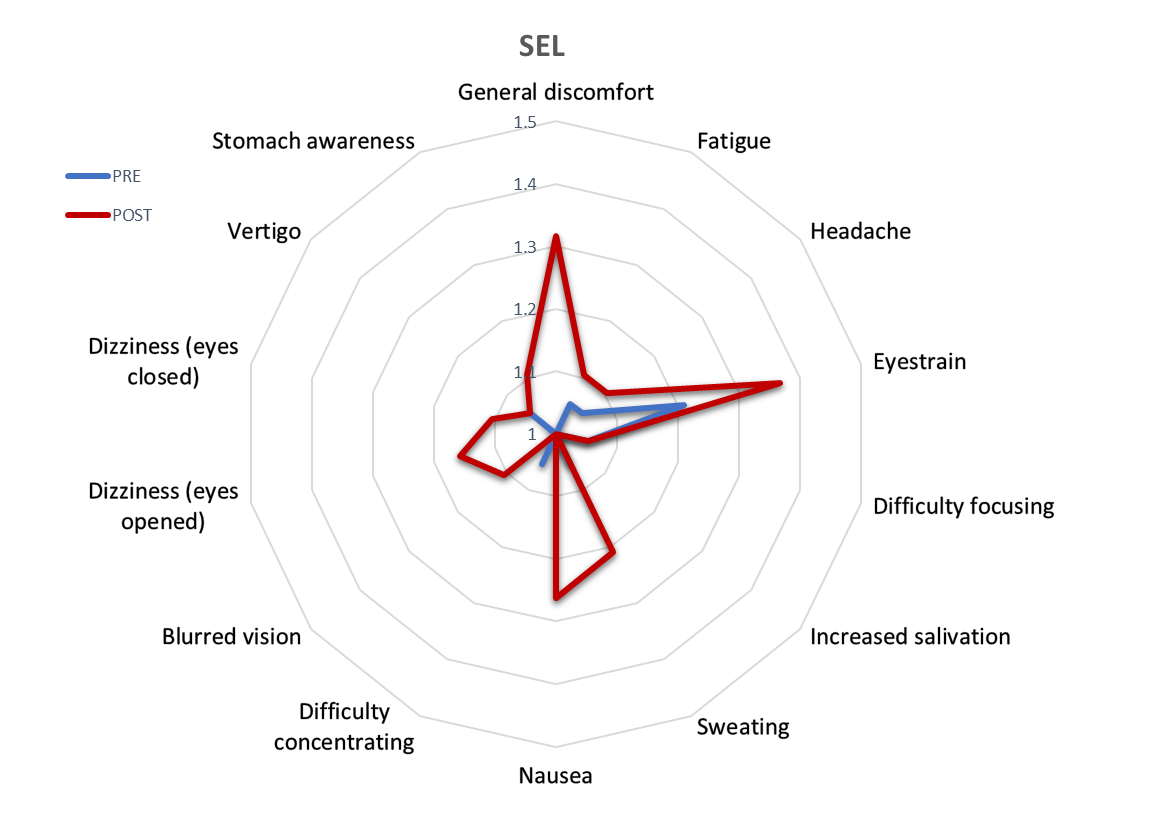}}
	\caption{Pre and post motion sickness questionnaire for the OMN (a) and SEL (b) interfaces. Health status is collected
before and after the experience using the Simulator Sickness
Questionnaire (SSQ) \cite{kennedy1993simulator}. Scale ranges from 1 to 4, but except for one case all ratings were equal or lower than 2. }
	\label{fig:salute2}
\end{figure}

In this Appendix, complete questionnaire results are provided. In particular, in Table \ref{tab:subjects}, characteristics of subjects involved in the user study are given (per group). Results concerning motion sickness are reported in Fig. \ref{fig:salute2}. Feedback collected through the other sections of the questionnaire are provided in Tables \ref{tab:questionnaire_UX}--\ref{tab:questionnaire_VR}. Specifically, subjective evaluation pertaining to questionnaire sections \textit{System competence}, \textit{Cognitive Load} and \textit{Overall User Experience} are reported in Table \ref{tab:questionnaire_UX}.  Subjective measurements for questionnaire section \textit{Reaction to test events} are reported in Table \ref{tab:questionnaire_testevents}, and complete the results presented in Fig. 6 of the main text. Subjective evaluation pertaining to the questionnaire section on \textit{Situational Awareness} are reported in Table \ref{tab:questionnaire_HMI}. Quality and quantity were evaluated for each element of the HMI, e.g., bounding boxes, navigation lines, etc., where quality in this context refers to how useful the specific element was in understanding the vehicle's behaviour as well as the surrounding environment. Finally, questions related to the VR experience (questionnaire section \textit{Immersion and presence}) are reported in Table \ref{tab:questionnaire_VR}. All questions were in Italian and had to be rated on a 1--5 Likert scale. The Selective (SEL) and Omni-comprehensive (OMN) AR-HUD interfaces are compared using the Mann-Whitney U-test. A $p$-value of .05 or lower was
considered to indicate a statistically significant difference; statistically significant differences are highlighted in bold in the tables.  

\begin{table}[h!]
	\centering 
	\caption{Subjects characteristics by study group}
	\begin{tabular}{clclllllclllll}
		\hline
		\multicolumn{2}{c}{\textbf{Test group}}    & \multicolumn{6}{c}{\textbf{Gender}}  & \multicolumn{6}{c}{\textbf{Age}}    \\ \hline
		\multicolumn{2}{c}{\multirow{2}{*}{SEL}} & \multicolumn{3}{c}{Male} & \multicolumn{3}{c}{Female} & \multicolumn{6}{c}{\multirow{2}{*}{22.53 $\pm$ 2.97}} \\ \cline{3-8}
		\multicolumn{2}{c}{}  & \multicolumn{3}{c}{12}   & \multicolumn{3}{c}{7} & \multicolumn{6}{c}{} \\ \hline
		\multicolumn{2}{c}{\multirow{2}{*}{OMN}} & \multicolumn{3}{l}{Male} & \multicolumn{3}{l}{Female} & \multicolumn{6}{c}{\multirow{2}{*}{ 25.27	$\pm$ 6.40}} \\ \cline{3-8}
		\multicolumn{2}{c}{}  & \multicolumn{3}{c}{13}   & \multicolumn{3}{c}{6} & \multicolumn{6}{c}{} \\ \hline
	\end{tabular}
    \label{tab:subjects}
\end{table}

\begin{table*}[h]
\caption{Subjective results on system competence, cognitive load, and user experience collected through questionnaires on a 1-to-5 (Disagree to agree) scale. Inspired by standard questions for the evaluation of trust in human-robot interaction (HRI),  this section evaluates the perceived AV's competence across the range of driving situations explored in the simulation . Overall user experience investigates general aspects regarding the mental model, and the trust posed in the system. }
\label{tab:questionnaire_UX}
\centering
\begin{tabular}{p{11cm}lllllll} 
\hline\hline
      & \multicolumn{2}{c}{SEL} &  & \multicolumn{2}{c}{OMN} &  & U-Test   \\ 
\cline{2-3}\cline{5-6}\cline{8-8}
\textbf{Statement} & \multicolumn{1}{c}{$\mu$}     & \multicolumn{1}{c}{$\sigma$}  &  & \multicolumn{1}{c}{$\mu$}     & \multicolumn{1}{c}{$\sigma$} &  & p-value  \\ 
\hline
      &  &   &  &  &   &  &     \\
\textit{\textbf{System Competence}}  &  &   &  &  &   &  &     \\
The autonomous vehicle showed adequate decision-making skills & 4.526 & .612  &  & 4.684 & .478  &  & .556     \\
The autonomous vehicle faced difficulties during unexpected changes in the environment   & 1.684 & .749  &  & 1.211 & .419  &  & .041     \\
The autonomous vehicle was smart     & 4.579 & .692  &  & 4.368 & .761  &  & .415     \\
I appreciated the driving skills of the autonomous vehicle    & 4.158 & .602  &  & 4.105 & .567  &  & .863     \\
      &  &   &  &  &   &  &     \\
\textit{\textbf{Cognitive Load}}     &  &   &  &  &   &  &     \\
It was demanding to find information within the HUD    & 1.000 & .000  &  & 1.053 & .229  &  & 1.000    \\
It was stressful to find information within the HUD    & 1.000 & .000  &  & 1.105 & .315  &  & .486     \\
Generally the amount of information provided by then HUD was excessive  & 1.053 & .809  &  & 2.105 & .229  &  & \textbf{.000   }  \\
Generally the comprehensibility of information provided by the HUD was adequate  & 4.737 & .562  &  & 4.684 & .582  &  & .908     \\
      &  &   &  &  &   &  &     \\
\textit{\textbf{Overall User Experience }}    &  &   &  &  &   &  &     \\
HUD information was useful to build trust in the vehicle  & 4.158 & .688  &  & 4.842 & .375  &  & .001     \\
HUD information was helpful to understand why the vehicle made a decision   & 4.263 & 1.046      &  & 4.842 & .375  &  & .055     \\
HUD information helped me feel comfortable and at ease    & 3.789 & 1.084      &  & 4.684 & .582  &  &\textbf{.003   }  \\
Thanks to HUD information, I had the perception that the vehicle was in full control of the situation     & 3.789 & 1.084      &  & 4.842 & .375  &  &\textbf{.000  }   \\
The HUD user interface was able to inform me before the potential danger affected driving & 2.421 & .607  &  & 4.105 & .567  &  & \textbf{.000}     \\
Generally, I have found that HUD information was helpful in anticipating the dangerous situation  & 2.474 & .612  &  & 4.421 & .607  &  & \textbf{.000 }    \\
   \\
\hline\hline
\end{tabular}
\end{table*}

\begin{table*}
\centering

\caption{Subjective assessment of test events collected through questionnaires on a 1-to-5 (Disagree to agree) scale. These questions provide complementary information to the physiological signals by investigating the perception of the event itself (sense of danger) as well as the quality of the HMI in presenting the event to the user (sense of surprise).}
\label{tab:questionnaire_testevents}
\begin{tabular}{p{11cm}lllllll} 
\hline\hline
    & \multicolumn{2}{l}{SEL} &  & \multicolumn{2}{l}{OMN} &  & U-Test   \\ 
\cline{2-3}\cline{5-6}\cline{8-8}
\textbf{Statement} & \multicolumn{1}{c}{$\mu$}     & \multicolumn{1}{c}{$\sigma$}  &  & \multicolumn{1}{c}{$\mu$}     & \multicolumn{1}{c}{$\sigma$} &  & p-value  \\ 
\hline
\\
\textit{\textbf{\textit{\textbf{Test Event  - }}Dog}} &  &   &  &  &   &  &     \\
How dangerous would you rate this situation? (not at all - very dangerous)    & 3.368 & .831  &  & 3.053 & .911  &  & .269     \\
I was surprised by this situation   & 4.421 & .769  &  & 3.737 & .991  &  &\textbf{.016     }\\
I detected the potential danger before it affected driving   & 1.632 & 1.065      &  & 2.737 & 1.046      &  &\textbf{.001}     \\
The information displayed was useful to anticipate the potential danger & 1.579 & .692  &  & 3.000 & .816  &  & \textbf{.001  }   \\
\textit{\textbf{\textit{\textbf{\textit{\textbf{Test Event  - }}}}Ball}}   &  &   &  &  &   &  &     \\
How dangerous would you rate this situation? (not at all - very dangerous)    & 3.053 & 1.079      &  & 3.000 & .943  &  & .872     \\
I was surprised by this situation   & 3.632 & 1.012      &  & 3.053 & .970  &  & .102     \\
I detected the potential danger before it affected driving   & 2.053 & 1.224      &  & 2.789 & 1.228      &  & .064     \\
The information displayed was useful to anticipate the potential danger & 1.579 & .692  &  & 3.263 & 1.368      &  &\textbf{.000   }  \\
\textit{\textbf{\textit{\textbf{\textit{\textbf{Test Event  - }}}}Car1}}   &  &   &  &  &   &  &     \\
How dangerous would you rate this situation? (not at all - very dangerous)    & 3.579 & .961  &  & 2.421 & 1.170      &  &\textbf{.003}     \\
I was surprised by this situation   & 3.368 & 1.300      &  & 1.947 & 1.129      &  & \textbf{.001 }    \\
I detected the potential danger before it affected driving   & 3.000 & 1.528      &  & 4.158 & 1.259      &  &\textbf{.007}     \\
The information displayed was useful to anticipate the potential danger & 2.632 & 1.012      &  & 3.947 & 1.129      &  & \textbf{.001 }    \\
\textit{\textbf{\textit{\textbf{Test Event  - }}Scooter}}    &  &   &  &  &   &  &     \\
How dangerous would you rate this situation? (not at all - very dangerous)    & 1.895 & 1.243      &  & 1.368 & .684  &  & .208     \\
I was surprised by this situation   & 1.842 & 1.068      &  & 1.684 & 1.003      &  & .628     \\
I detected the potential danger before it affected driving   & 2.105 & 1.286      &  & 2.000 & 1.291      &  & .773     \\
The information displayed was useful to anticipate the potential danger & 1.211 & .713  &  & 1.053 & .229  &  & .743     \\
\textit{\textbf{\textit{\textbf{Test Event  - }}Car2}}     &  &   &  &  &   &  &     \\
How dangerous would you rate this situation? (not at all - very dangerous)    & 4.211 & .787  &  & 3.421 & 1.017      &  &\textbf{.017  }   \\
I was surprised by this situation   & 4.105 & .994  &  & 3.053 & .970  &  & .002     \\
I detected the potential danger before it affected driving   & 1.895 & .994  &  & 3.316 & 1.204      &  &\textbf{.000  }   \\
The information displayed was useful to anticipate the potential danger & 2.053 & .848  &  & 3.158 & .958  &  &\textbf{.000}   \\
\textit{\textbf{\textit{\textbf{Test Event  - }}Man1}}     &  &   &  &  &   &  &     \\
How dangerous would you rate this situation? (not at all - very dangerous)     & 1.316 & .671  &  & 1.263 & .452  &  & 1.000    \\
I was surprised by this situation   & 1.316 & .582  &  & 1.158 & .501  &  & .390     \\
I detected the potential danger before it affected driving   & 3.474 & 1.712      &  & 4.474 & .772  &  & .093     \\
The information displayed was useful to anticipate the potential danger & 3.421 & 1.170      &  & 4.316 & .885  &  &\textbf{.018  }   \\
\textit{\textbf{\textit{\textbf{\textit{\textbf{Test Event  - }}}}Man2}}   &  &   &  &  &   &  &     \\
How dangerous would you rate this situation? (not at all - very dangerous)    & 4.474 & .697  &  & 3.737 & .933  &  &\textbf{.008  }   \\
I was surprised by this situation   & 4.526 & .841  &  & 3.368 & 1.012      &  & \textbf{.000}     \\
I detected the potential danger before it affected driving   & 1.842 & 1.015      &  & 3.053 & 1.079      &  &\textbf{.001 }    \\
The information displayed was useful to anticipate the potential danger & 1.474 & .697  &  & 3.158 & .958  &  & \textbf{.001}     \\
    &  &   &  &  &   &  &     \\\hline\hline
\end{tabular}
\end{table*}

\begin{table*}
\centering

\caption{Subjective assessment of HMI elements collected through questionnaires on a 1-to-5 (Disagree to agree) scale. Selective (SEL) and Omni-Comprehensive (OMN) AR-HUD interfaces are compared using the Mann-Whitney U-test}
\label{tab:questionnaire_HMI}
\begin{tabular}{p{11cm}lllllll} 
\hline\hline
    & \multicolumn{2}{l}{SEL} &  & \multicolumn{2}{l}{OMN} &  & U-Test   \\ 
\cline{2-3}\cline{5-6}\cline{8-8}
\textbf{Statement} & \multicolumn{1}{c}{$\mu$}     & \multicolumn{1}{c}{$\sigma$}  &  & \multicolumn{1}{c}{$\mu$}     & \multicolumn{1}{c}{$\sigma$} &  & p-value  \\ 
\hline
\\
\textit{\textbf{Situational Awareness}}     &  &   &  &  &   &  &     \\
Bounding boxes (BB) helped me understand that the car had taken charge of the traffic lights and handled them appropriately   & 4.824 & .393  &  & 4.765 & .437  &  & 1.000    \\
Labels helped me understand that the car had taken charge of the traffic lights and handled them appropriately   & 4.842 & .501  &  & 4.947 & .229  &  & .743     \\
BB helped me understand that the car had taken charge of the road sign and handled it appropriately  & 4.556 & .784  &  & 4.500 & .730  &  & .772     \\
Labels helped me understand that the car had taken charge of the road sign and handled it appropriately & 4.684 & .478  &  & 4.684 & .671  &  & .714     \\
BB helped me understand that the car had taken charge of the potential obstacle (pedestrian, animal) and handled it appropriately & 4.316 & 1.157      &  & 4.737 & .452  &  & .492     \\
Labels helped me understand that the car had taken charge of the potential obstacle (pedestrian, animal) and handled it appropriately  & 4.263 & 1.195      &  & 4.579 & .838  &  & .558     \\
BB helped me to understand that the car had taken charge of the other cars likely affecting the driving and figured out how to handle them      & 4.895 & .315  &  & 4.895 & .315  &  & 1.000    \\
Labels helped me to understand that the car had taken charge of the other cars likely affecting the driving and figured out how to handle them  & 4.842 & .501  &  & 4.684 & .749  &  & .500     \\
Navigation line has been helpful in understanding the vehicles's intentions  & 4.944 & .236  &  & 4.947 & .229  &  & 1.000    \\
Other vehicles' navigation lines helped me to understand that the car had taken charge of their proximity and figured out how to handle them   & 5.000 & .000  &  & 4.579 & .961  &  & \textbf{.046  }   \\
BB colour linked to the level of risk helped me to understand that the car had taken charge of obstacles and handled them appropriately& 4.800 & .561  &  & 4.600 & .737  &  & .555     \\
The warning sound of traffic light/road sign helped me to understand that the car had taken charge of the situation     & 4.688 & .602  &  & 4.250 & .775  &  & .106     \\
The warning sound in case of danger helped me to understand that the car had taken charge of the situation and handled it appropriately    & 4.500 & .857  &  & 4.588 & .618  &  & 1.000    \\
    &  &   &  &  &   &  &     \\
\textit{\textbf{Quantity/Mental Workload: how would you rate the quantity of information? (1=poor, 3=adequate, 5=excessive)}}   &  &   &  &  &   &  &     \\
Number of bounding boxes and labels for traffic lights  & 3.053 & .229  &  & 3.105 & .459  &  & .604     \\
Number of bounding boxes and labels for road signs & 2.579 & .607  &  & 3.263 & .562  &  &\textbf{.001 }    \\
Number of bounding boxes and labels for potential obstacles (pedestrian, animal, etc.)     & 2.474 & .697  &  & 3.105 & .315  &  & \textbf{.001}     \\
Number of bounding boxes and labels for traffic cars    & 2.789 & .419  &  & 3.474 & .697  &  &\textbf{.001   }  \\
Number of navigation lines for the traffic cars    & 2.842 & .375  &  & 3.474 & .905  &  &\textbf{.014  }   \\
The warning sound of traffic light/road sign  & 2.368 & .684  &  & 3.316 & .582  &  &\textbf{.000 }    \\
The warning sound in case of danger    & 2.526 & .697  &  & 3.000 & .000  &  & \textbf{.008   }  \\
    &  &   &  &  &   &  &     \\
\hline\hline
    &  &   &  &  &   &  &    
\end{tabular}
\end{table*}

\begin{table*}
\centering
\caption{Subjective assessment of VR simulator collected through questionnaires on a 1-to-5 (Disagree to agree) scale. Relevant sections from the VRUSE questionnaire selected to evaluate the simulated environment with respect to immersion, presence and fidelity.}
\label{tab:questionnaire_VR}
\begin{tabular}{p{11cm}lllllll} 
\hline\hline
      & \multicolumn{2}{l}{SEL} &  & \multicolumn{2}{l}{OMN} &  & U-Test   \\ 
\cline{2-3}\cline{5-6}\cline{8-8}
\textbf{Statement} & \multicolumn{1}{c}{$\mu$}     & \multicolumn{1}{c}{$\sigma$}  &  & \multicolumn{1}{c}{$\mu$}     & \multicolumn{1}{c}{$\sigma$} &  & p-value  \\ 
\hline
      &  &   &  &  &   &  &     \\
\textit{\textbf{Immersion and Presence}}    &  &   &  &  &   &  &     \\
I felt a sense of being immersed in the virtual environment  & 4.526 & .697  &  & 4.474 & .772  &  & .999     \\
The quality of the image reduced my feeling of presence   & 2.105 & 1.150      &  & 2.474 & 1.020      &  & .239     \\
I had a good sense of scale in the virtual environment    & 4.947 & .229  &  & 4.842 & .375  &  & .604     \\
The presence of my hands and legs within the VR enhanced my sense of presence  & 4.632 & .684  &  & 4.579 & .769  &  & 1.000    \\
The motion platform enhanced my sense of presence    & 4.474 & 1.073      &  & 4.579 & .838  &  & .987     \\
Overall I would rate my sense of presence as: (1) very unsatisfactory- (5) very satisfactory    & 4.368 & .597  &  & 4.421 & .692  &  & .769     \\
      &  &   &  &  &   &  &     \\
\textit{\textbf{ Simulation Fidelity}}  &  &   &  &  &   &  &     \\
The virtual driving simulation was accurate & 4.368 & .895  &  & 4.368 & .496  &  & .571     \\
Objects in the virtual environment moved in a natural way & 4.158 & .958  &  & 4.053 & .621  &  & .359     \\
The simulation seemed to stop at times   & 1.000 & .000  &  & 1.000 & .000  &  & 1.000    \\
I had a realistic perception about the virtual objects    & 4.105 & .737  &  & 3.789 & .855  &  & .315     \\
The experience in the virtual world was consistent with what I could have lived in the real world    & 4.421 & .692  &  & 4.368 & .761  &  & .966     \\
Movements of the motion platform were realistic  & 4.053 & 1.129      &  & 4.158 & .898  &  & .994     \\
Overall I would rate the faithfulness of the driving simulation as: (1) very unsatisfactory- (5) very satisfactory    & 4.316 & .582  &  & 4.421 & .607  &  & .662     \\ 
\hline\hline
      &  &   &  &  &   &  &    
\end{tabular}
\end{table*}

\end{document}